\documentclass[11pt,a4paper]{article}
\usepackage{jinstpub}
\usepackage{amsfonts}
\let\savebigtimes\bigtimes
\let\bigtimes\relax
\usepackage{mathabx}
\let\bigtimes\savebigtimes
\usepackage{lmodern}
\usepackage{siunitx}
\DeclareSIUnit\gauss{G}
\usepackage{physics}
\usepackage[hang]{footmisc}
\usepackage{enumitem}
\usepackage{adjustbox}
\usepackage[all]{nowidow}
\usepackage{rotating}
\usepackage{pdflscape}
\usepackage{multirow}
\usepackage{booktabs}
\usepackage[font=small,bf,labelsep=period]{caption}
\usepackage{microtype}
\usepackage{graphicx,color}
\usepackage{hyperref}
\usepackage{doi}
\usepackage{CJKutf8}
\usepackage{etoolbox}
\usepackage{units}
\usepackage{physics}

\usepackage{subcaption}

\title{Novel Light Field Imaging Device with Enhanced Light Collection for Cold Atom Clouds}
\author[a,b]{S.~Cheong,}
\author[b]{J.C.~Frisch,}
\author[b]{S.~Gasiorowski,}
\author[a]{J.M.~Hogan,}
\author[b]{M.~Kagan,}
\author[a,b,1]{M.~Safdari%
\note{Corresponding author.},}
\author[b]{A.~Schwartzman,}
\author[b]{and M.~Vandegar}
\affiliation[a]{Stanford University, 450 Serra Mall, Stanford, CA 94305, USA}
\affiliation[b]{SLAC National Accelerator Laboratory, 2575 Sand Hill Road, Menlo Park, CA 94025, USA}
\emailAdd{murtazas@stanford.edu}
\abstract{We present a light field imaging system that captures multiple views of an object with a single shot. The system is designed to maximize the total light collection by accepting a larger solid angle of light than a conventional lens with equivalent depth of field. This is achieved by populating a plane of virtual objects using mirrors and fully utilizing the available field of view and depth of field. Simulation results demonstrate that this design is capable of single-shot tomography of objects of size $\mathcal{O}$(1~mm$^3$), reconstructing the 3-dimensional (3D) distribution and features not accessible from any single view angle in isolation. In particular, for atom clouds used in atom interferometry experiments, the system can reconstruct 3D fringe patterns with size $\mathcal{O}$(100~µm). We also demonstrate this system with a 3D-printed prototype. The prototype is used to take images of $\mathcal{O}$(1~mm$^{3}$) sized objects, and 3D reconstruction algorithms running on a single-shot image successfully reconstruct $\mathcal{O}$(100~µm) internal features. The prototype also shows that the system can be built with 3D printing technology and hence can be deployed quickly and cost-effectively in experiments with needs for enhanced light collection or 3D reconstruction. Imaging of cold atom clouds in atom interferometry is a key application of this new type of imaging device where enhanced light collection, high depth of field, and 3D tomographic reconstruction can provide new handles to characterize the atom clouds.}

\begin{document}

\maketitle
\flushbottom
\newpage


\section{Introduction}
\label{sec:intro}

We describe a new type of system for imaging a 3-dimensional object in low light conditions, for example cold atom clouds in atomic sensors. The system operates by using an array of mirrors to project multiple views of the object onto a single high spatial resolution imager, thus measuring both the intensity and direction of emitted light, e.g. the light field. This allows collection of photons over a large solid angle, while maintaining a high $f$-number for a large depth of field. 

A key application of this novel device is imaging cold atom clouds in atomic sensors, such as atom interferometers. Atomic sensors offer an exciting new opportunity to expand the reach and scope of High Energy Physics. In particular, long-baseline atom interferometer gradiometers can provide sensitivity to ultralight dark matter candidates with a mass between $\unit[10^{-16}]{eV}$ and $\unit[10^{-12}]{eV}$, well below what current and planned detection techniques could reach~\cite{Abe_2021}. Scaling up the size of these quantum sensors will also open a new window in the exploration of gravitational waves (GW) in the mid-band ($\unit[0.03]{Hz}$ -- $\unit[3]{Hz}$) range.

Achieving the full potential of these type of future atomic sensors will require new and advanced methods for imaging atom clouds that are capable of capturing as much light as possible while maintaining high depth of field, as well as providing the ability to image the atom cloud from multiple angles for tomographic reconstruction.





Maximizing light collection is important to reduce the fluorescence imaging time. The longer the atoms are illuminated for imaging purposes, the more the atom cloud distorts due to heating). At the same time, the size of the atom cloud sets a minimum requirement for depth of field. Conventional imaging systems consisting of a lens and sensor have a fundamental limit for depth of field vs numerical aperture that do not allow maximization of the first two requirements simultaneously. On the one hand, maximizing light collection calls for a large numerical aperture lens. On the other hand, the large depth of field requirement (of the order of the $\unit[1]{mm}$ atom cloud), calls for a low aperture lens. The ability to capture images from multiple angles would enable the characterization of key sources of systematic uncertainty, for example for measurements of laser wavefront aberration phase errors~\cite{Abe_2021} via 3D tomographic reconstruction.


In this paper, we propose a new type of imaging device that can meet these three key needs for enhanced image and physics precision in atom intereformetry. We focus on the application for the MAGIS-100 Experiment, currently under construction at Fermilab, but the approach is general and could be applied to any scientific (or industrial) application that has similar constraints in terms of achieving high depth of field imaging with low illumination conditions and/or 3-dimensional imaging capabilities.

\section{Imaging requirements in atom interferometry}
\label{sec:imaging_req}



Atom interferometers rely on ensembles of cold atoms that can be switched between stable internal states using pulses of light.  The light pulse sequence also serves to transfer momentum to the atoms, splitting the atomic wavefunction into two paths that subsequently interfere, realizing a Mach-Zehnder interferometer geometry.  MAGIS-100 will rely on Sr atoms, and with typical operating parameters is expected to produce clouds of $\mathcal{O}(10^6)$ Sr atoms in a Gaussian density profile within a volume of $\unit[1]{mm^3}$. At the end of the interferometer pulse sequence, the final interference pattern consists of two output ports, one for each of the final interfering atomic states.  Each atom from the initial ensemble lands in one of the two ports with a probability determined by the phase accumulated during the interferometer.  By the application of a phase shear readout (PSR) technique~\cite{PhysRevLett.111.113002}, a spatial fringe pattern is imprinted on the two output ports. By unitarity, the resulting fringes that appear on the two interferometer ports are offset by 180 degrees with respect to one another. MAGIS-100 is expected to produce PSR interference fringes with a wavelength of $\mathcal{O}(\unit[100]{\mu m})$ over the $\unit[1]{mm^3}$ atom clouds.


The phase of the interference pattern that is imprinted on the atom cloud density is a key physics observable in the MAGIS-100 experiment. This drives the minimum feature size of the imaging systems used to capture the fluorescence light from the atom clouds in service of finding the interference phase. Each of the Sr atoms at MAGIS-100 will scatter $10^{8}$ photons per second when excited by the imaging light pulse at $\unit[461]{nm}$. The imaging light pulse will be turned on for a duration between $\unit[10]{\mu s}$ -- $\unit[100]{\mu s}$. Here the upper limit is set by the requirement that diffusive heating from the laser does not average out the interference pattern.

The above sets the minimal requirements and parameters for any imaging system that will be deployed to image the atom interferometer fringe patterns at MAGIS-100. Focusing on imaging a single atom cloud, these are listed below:
\begin{itemize}
  \item Minimum resolution capable of distinguishing $\mathcal{O}(\unit[100]{\mu m})$ features.
  \item Field of view and depth of field to accommodate the $\mathcal{O}(\unit[1]{mm^3})$ atom cloud envelope
  \item Maximize the light collected from the atom cloud to minimize imaging time and reduce photon shot noise. 
\end{itemize}

\section{Optical design}
\label{sec:optical}

\subsection{Conventional imaging considerations and limitations}
\label{sec:conventional_imager}
A conventional imaging system can be described as camera and a finite conjugate lens focused on the object of interest. Following the imaging requirements in the MAGIS-100 experiment as described in section~\ref{sec:imaging_req}, a simple conventional imaging system can be designed from the geometric model of image formation (reviewed with relevant details in appendix \ref{appendix:entocentric}). For this study a CMOS sensor with a pixel size of 3 $\mu$m is considered with a total sensor size that is arbitrarily large such that it can accommodate the full 1~mm$^3$ atom cloud image on the sensor. Therefore the problem reduces to a choice of appropriate parameters of the entocentric lens that is used in the conventional imaging of the atom clouds. 

\begin{figure}
    \centering
    \includegraphics[width=0.49\textwidth]{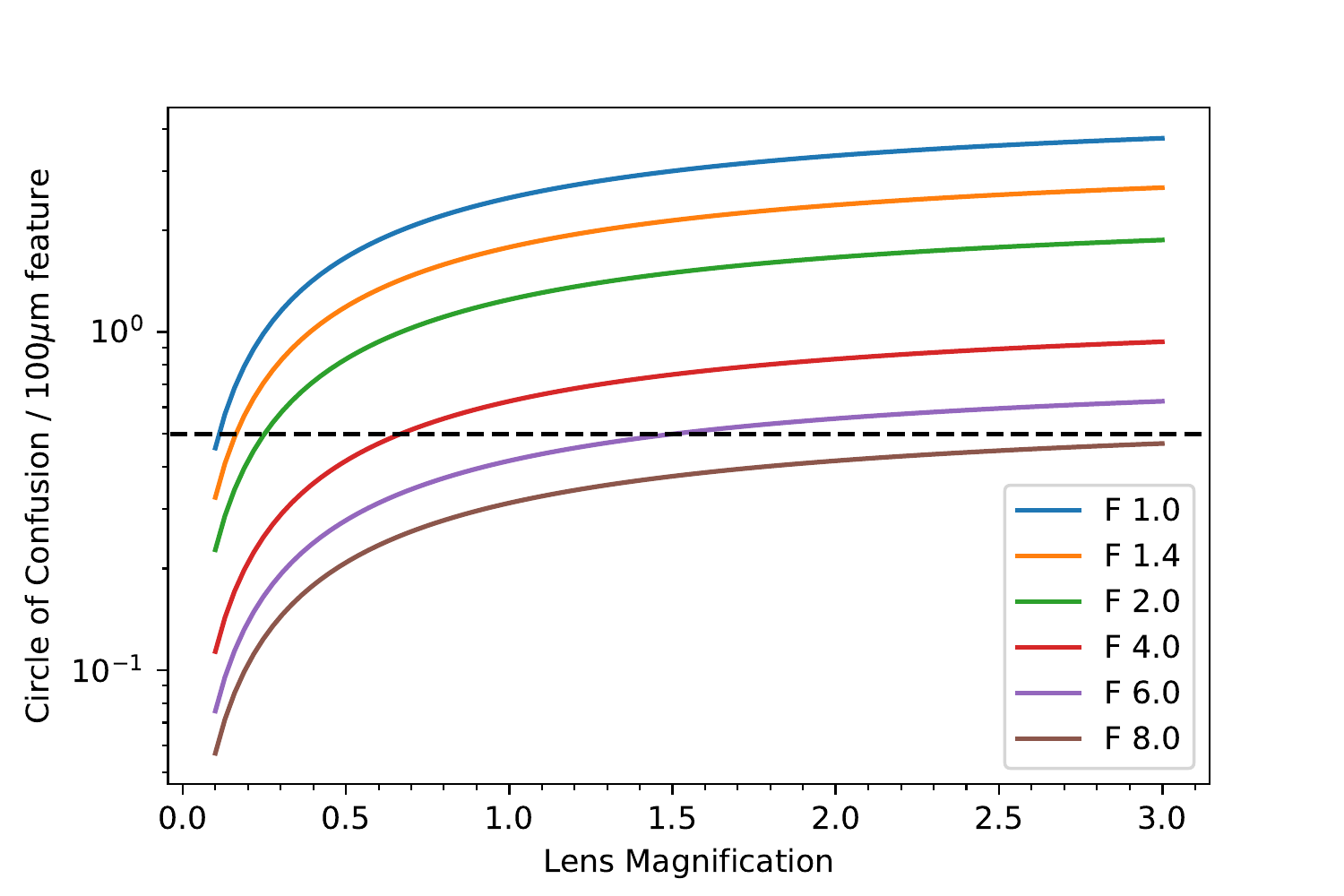}
    \includegraphics[width=0.49\textwidth]{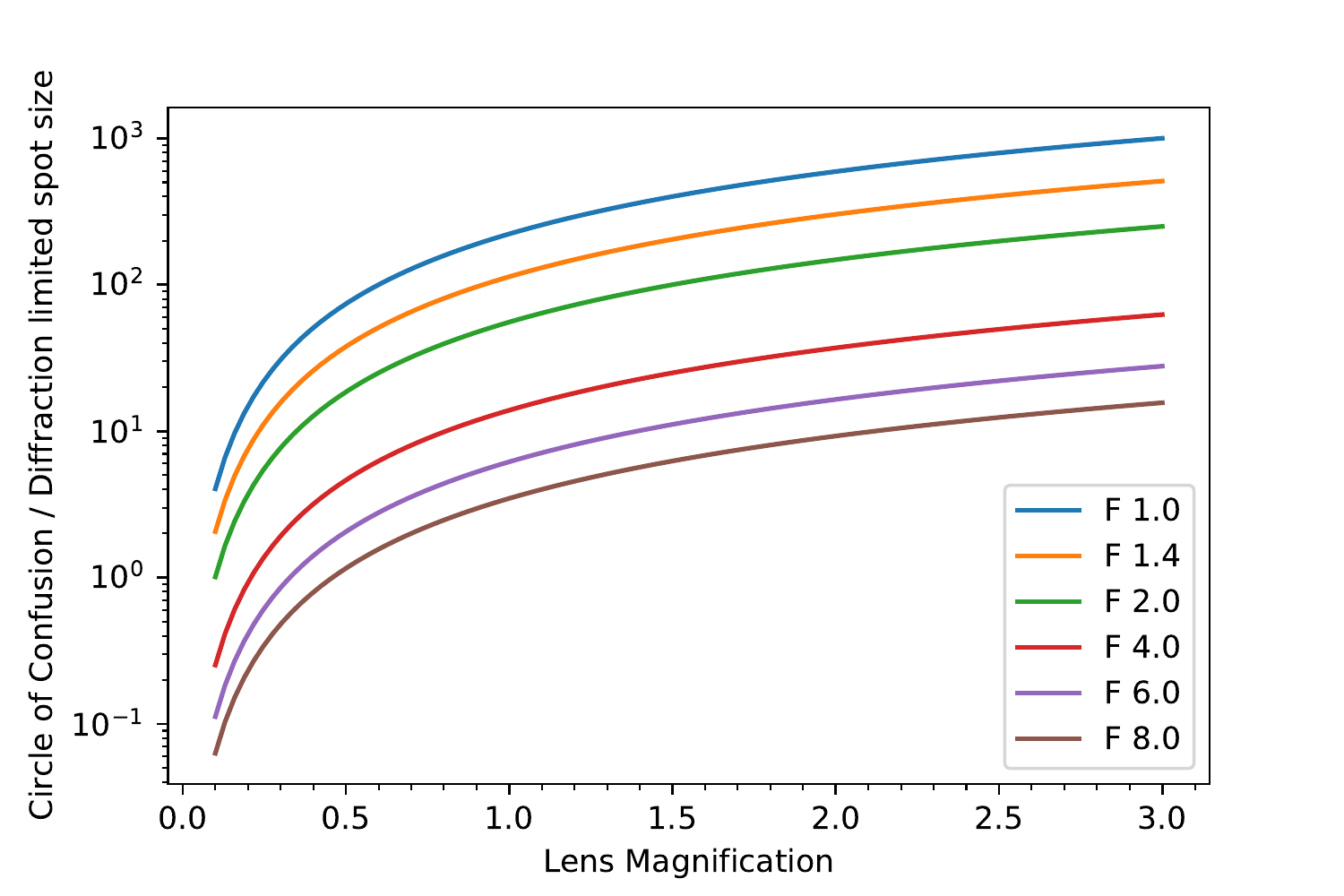}
    \caption{(Left) Circle of confusion in an ideal conventional single view entocentric imager relative to the size of a $\unit[100]{\mu m}$ feature, as a function of lens magnification for a range of lens $f$-numbers. The $\unit[100]{\mu m}$ object space feature dimensions are divided by the magnification to get the image space dimension, which is then used to divide the circle of confusion. The horizontal line is drawn for the points where the circle of confusion is 50\% of the $\unit[100]{\mu m}$ feature on the imaging plane. (Right) Circle of confusion in an ideal conventional single view entocentric imager relative to the diffraction limited spot size, as a function of lens magnification for a range of lens $f$-numbers. The circle of confusion sizes are computed using eq.~\ref{eqn:ento_dof} with a $f = \unit[62]{mm}$, though as seen from eq.~\ref{eqn:ento_dof_simplified} the dependence on the focal length is not a dominant effect. The depth of field is set to $\unit[1]{m}$ to account for the typical size of the cold atom clouds in atom interferometers like MAGIS-100.}
    \label{fig:lens_param_scan}
\end{figure}

A scan of lens magnifications $m$ and $f$-number $N$ is presented in figure~\ref{fig:lens_param_scan} where the impact of various combinations can be seen in terms of the resulting circle of confusion for point source imaging. The depth of field is set to be $\unit[1]{mm}$ per section~\ref{sec:imaging_req}. Note here that the circle of confusion size is not reported as a raw length or number of pixels, but rather as a fraction of the size a $\unit[100]{\mu m}$ feature takes on the imaging plane. This helps contextualize the size of the circle of confusion in terms of the minimum feature size the system is designed to capture. Figure~\ref{fig:lens_param_scan} also shows the size of the geometric circle of confusion relative to the diffraction limited spot size. Here the diffraction spot size is approximated using the Fraunhofer limit, i.e. the first zero of the Airy Disk ($2.44\lambda N$) \cite{hecht1987optics} where $N$ is the lens $f$-number and the wavelength $\lambda = \unit[461]{nm}$ is used which is the Sr imaging wavelength.

\begin{figure}
    \centering
    \includegraphics[width=0.49\textwidth]{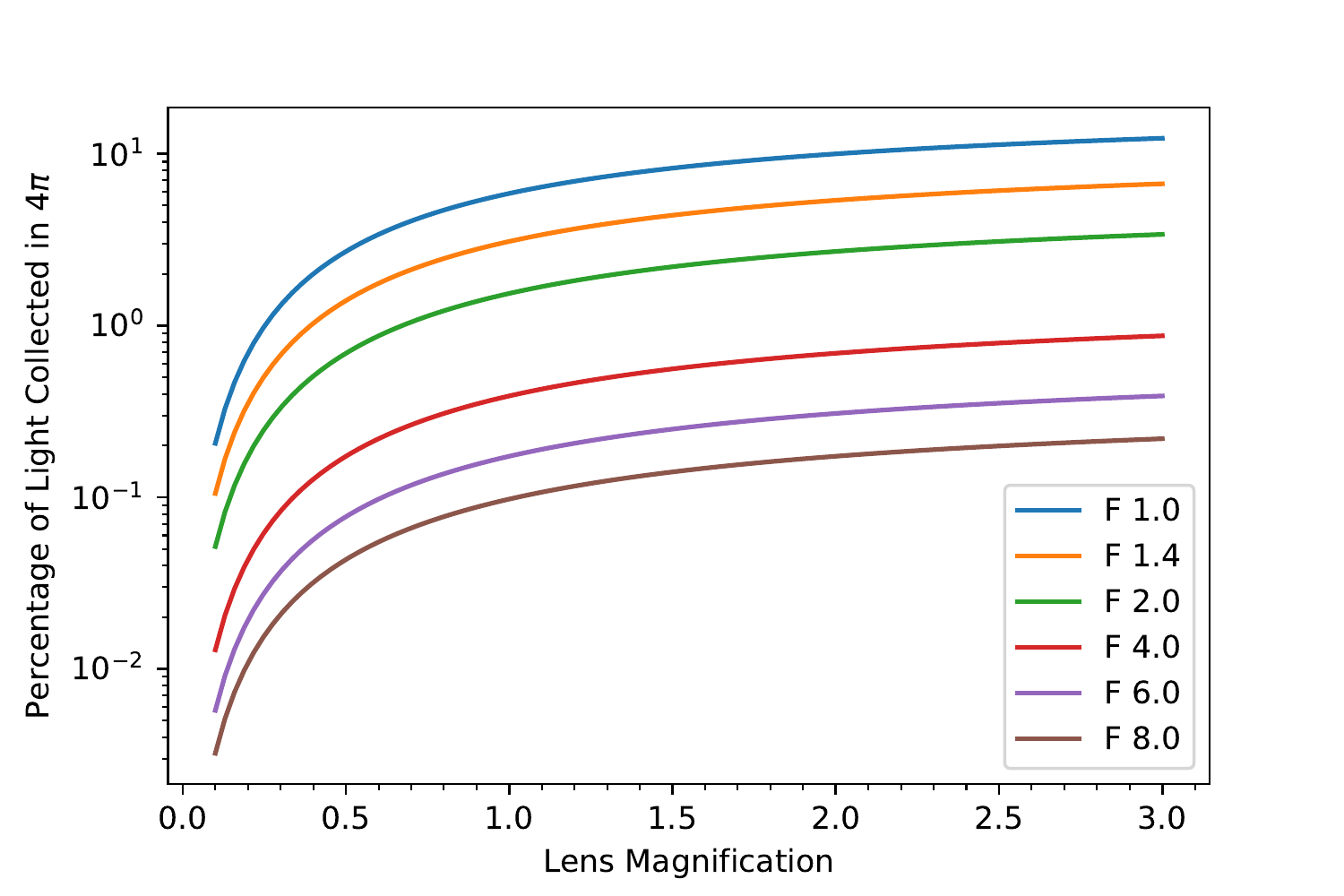}
    \caption{Percentage of light collected by an ideal conventional single view entocentric imager that is emanated by a point object in $4\pi$, as a function of lens magnification for a range of lens F-numbers. The percentages are computed using eq.~\ref{eqn:ento_light}.}
    \label{fig:lens_param_scan_light}
\end{figure}

Figure~\ref{fig:lens_param_scan} helps constrain the choice of lens parameters to a space where the circle of confusion is less than half the size of the $\unit[100]{\mu m}$ feature (below the horizontal line). Note that this is a pessimistic analysis that assumes the atom cloud density to be uniform in the full $\unit[1]{mm^3}$ volume, whereas the atom clouds are expected to have a Gaussian density profile. Nevertheless it helps isolate a part of the lens parameter space that can be exploited for high fidelity imaging of the atom clouds. The fraction of light collected from an isotropic point source by lenses in this space reaches a maximum of approximately 0.3\%, as shown in figure~\ref{fig:lens_param_scan_light}. 

This presents a challenge in utilizing conventional entocentric imaging for atom interferometry; prioritizing light collection jeopardizes the ability of the imaging system to resolve the features of the interference fringes and vice versa. However, the target object here being only $\unit[1]{mm^3}$ opens up the possibility to collect more light on the same sensor by utilizing the available pixels and folding in more optical paths into the same imaging system. Furthermore, collecting light from a larger set of angles around the object captures more of the light field, which in turn provides more information to reconstruct the object from the captured image. This is described in detail below.

\subsection{Light field imaging concept for atom interferometers}
\label{sec:lightfield_imager}


\subsubsection{Geometry of a single folded view}
\label{sec:lightfield_single_view}

\begin{figure}
    \centering
    \includegraphics[width=0.9\textwidth]{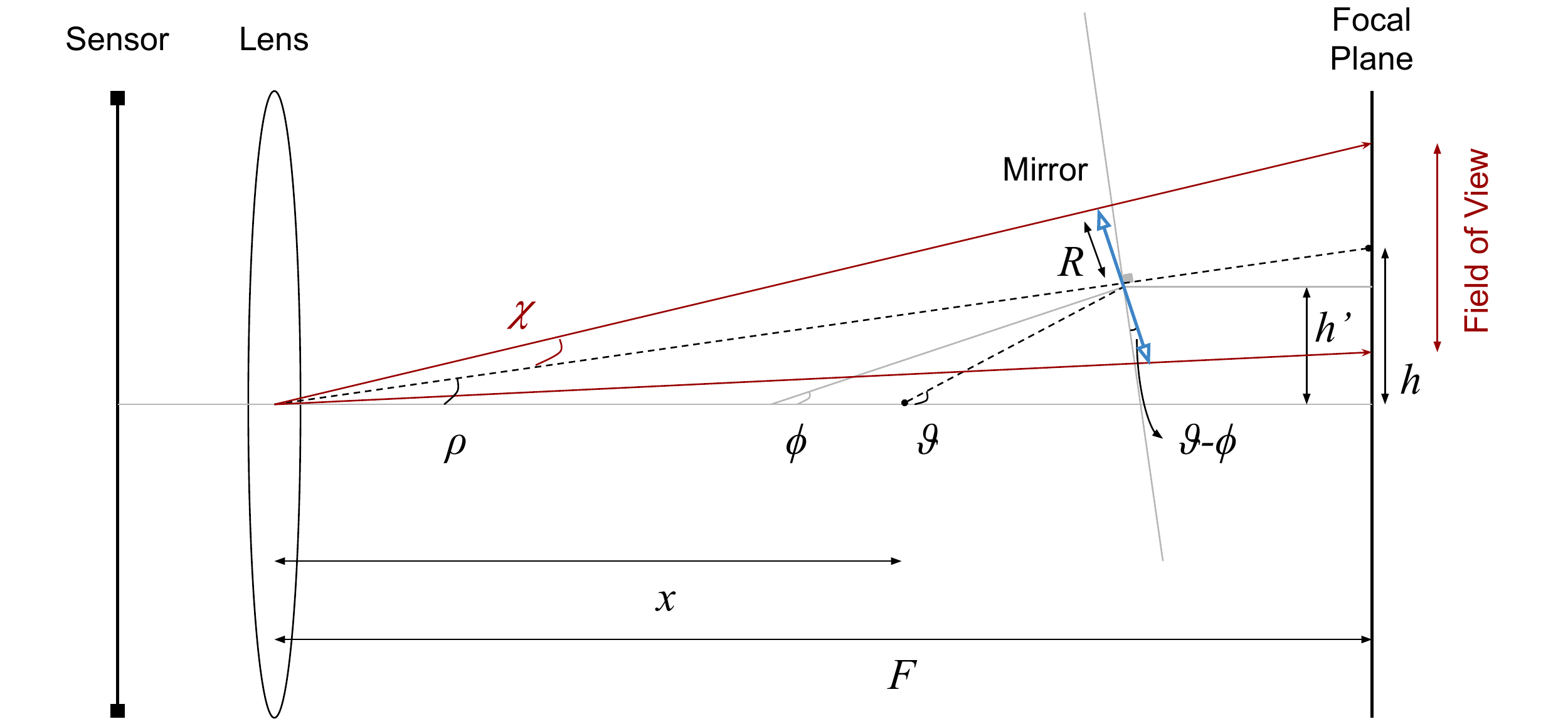}
    \caption{Ray diagram of a single folded view. Here the object is placed at a distance $x$ from the lens center, offset from the focal plane of the lens which is at a distance $F$ from the lens. A flat mirror is introduced between the object and the focal plane, which produces a virtual object that lies on the focal plane; this virtual object is focused onto the sensor by the lens. The mirror normal angle $\phi$ is chosen such that the principal (central) ray emanating from the object (at angle $\theta$) is reflected towards the center of the lens. This constrains the geometry of the mirror and determines the mirror normal angle as described in appendix \ref{appendix:normal_equations}. Note that the $\theta>0$ view angle results in an image that sees the object as if it were rotated by angle $\theta$ about the optical axis. Note also that $\rho = 2\phi - \theta$. The marginal rays from the lens center to the mirror edges are shown in red, and corresponding field of view is highlighted on the focal plane. The half angle $\chi$ corresponds to the angular field of view and is determined following the logic presented in appendix \ref{appendix:fov_equations}.}
    \label{fig:singleview_folded_fov}
\end{figure}

The concept of the light field imager presented here relies on folding optical paths using flat mirrors. Figure~\ref{fig:singleview_folded_fov} illustrates the key properties of this folding method for a single view.\footnote{This text uses the word "view" to refer to mirror folded views.} For a given lens and sensor, the focal plane of the imaging apparatus is some well defined surface which we can assume is flat for the case of an ideal lens. Rather than placing the object at the focal plane directly, the object is placed closer to the imaging apparatus, and a flat mirror is used to create a virtual copy of the object that sits at the focal plane of the lens. Note that the relative angle between the mirror and optical axis creates a virtual object that is rotated with respect to the imaging apparatus. Light from this virtual object is then focused onto the sensor, along with the light from the object directly that forms a defocused spot on the center of the sensor. Since the focused image on the sensor can be mapped to a specific view angle, this setup preserves both the intensity and angular information of the light emanating from the object --- capturing the light field.

The principal ray passing through the center of the lens is highlighted as the dashed line in figure~\ref{fig:singleview_folded_fov}. For a given view angle $\theta$, the constraints that the virtual object must sit on the focal plane, and that the principal ray must pass through the center of the lens, uniquely determine the mirror normal angle $\phi$, and the position of the mirror center. This is detailed more explicitly in appendix \ref{appendix:normal_equations}.

The point of reflection of the principal rays that pass through the lens center can be treated as the center of the mirror, and a symmetrically shaped mirror can ensure that the light cone emanating from the object reflecting off the mirror stays symmetric about the center. Note that the distance between the virtual object and lens increases by a factor of $1/\cos{\rho}$ as the virtual object rises above the optical axis. Furthermore, the perpendicular area of the lens aperture (as seen by the virtual object) decreases by a factor of $\cos{\rho}$. Therefore, when the lens is the light-limiting aperture, this leads to a drop in the light collected from the object as the view gets increasingly off-axis. This effect can be mitigated by using mirrors that function as the light-limiting apertures, and increasing the relevant mirror dimension by the appropriate factor of $1/\cos{(\theta-\phi)}$ for each mirror. Alternatively, if mirrors of fixed radii are used that form the light-limiting apertures, then a drop in light intensity collected at the lens is expected from similar factors of $\cos{(\theta-\phi)}$.

The field of view for a single folded view can be geometrically determined by tracing marginal rays from the center of the lens to the edges of the mirror, as shown in figure~\ref{fig:singleview_folded_fov}. Under the assumption that the mirror is symmetric about its center, this can be done following the details presented in appendix \ref{appendix:fov_equations}. The mirror sizes can be chosen based on the field of view to account for the object of interest. Therefore the system is flexible in its ability to accommodate objects of various scales.

\subsubsection{Accommodating multiple views -- Capturing the light field}
\label{sec:lightfield_multi_view}

\begin{figure}
    \centering
    \includegraphics[width=0.9\textwidth]{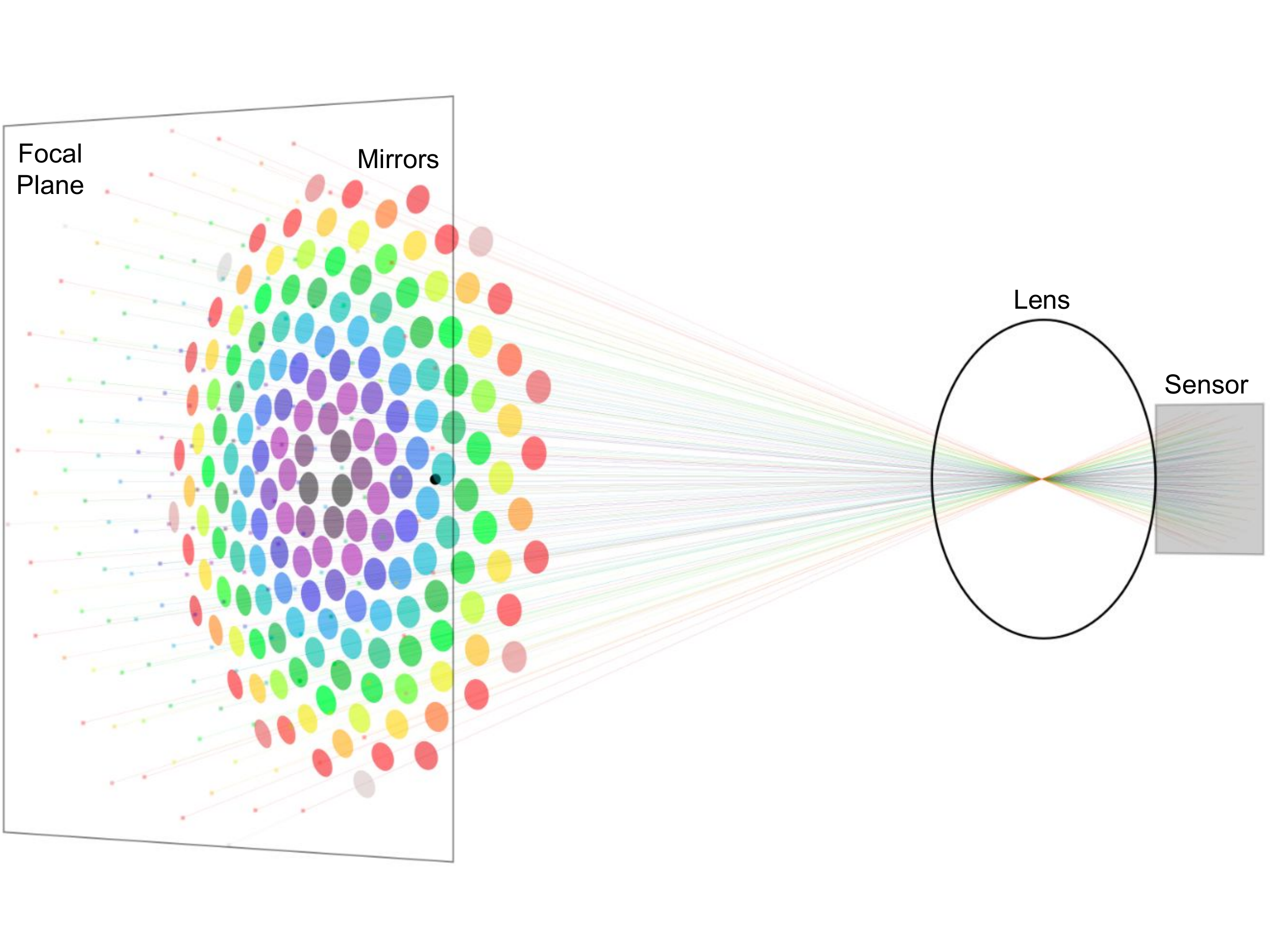}
    \caption{Principal ray diagram in a setup with multiple view folding using mirrors. The object is shown as a black dot along the optical axis of the system. Each mirror is colored from blue to red proportionally to the magnitude of the polar view angle $\theta$ captured by the mirror. The principal rays are traced with the same color, and are shown to start at the location of the virtual object and end at the sensor. The focal plane is depicted here with a bounding box that corresponds to the size of the sensor divided by the lens magnification. It can be seen that for the given array of mirrors, there is some minimum distance between neighboring virtual objects centers. The density of virtual objects on the focal plane can be modified by increasing or decreasing the number of views and mirrors to appropriately account for the virtual object size. Note here that the view angles $(\theta, \psi)$ are sampled on the Fibonacci lattice as governed by eq.~\ref{eqn:fibonacci}, but only those views (and corresponding mirrors) within the appropriate hemisphere, and which produce virtual objects that lie inside the field of view of the imaging apparatus, are shown.}
    \label{fig:multiview}
\end{figure}

The single view geometry can be readily extended to a multi view geometry given that the image of each virtual object occupies a small portion of the imaging sensor. This can be visualized by mapping the image sensor dimensions onto the focal plane of the lens, as shown in figure~\ref{fig:multiview}. In this setup, the view angles $(\theta, \psi)$ are first sampled on a Fibonacci lattice as follows,
\begin{align}
\label{eqn:fibonacci}(\theta_i, \psi_i) &= \left(\arccos{\left[ 1 - 2\frac{i}{N}\right] }, 2\pi \frac{i}{GR}\right)
\end{align}
where $GR = \frac{1 + \sqrt{5}}{2}$, $i=(1/2 + j)$ where $j$ is an integer from $0$ to $N$, and $N$ is the total number of points being sampled on the Fibonacci lattice. Here $\theta_i$ correspond to angles with respect to the optical axis, and $\psi_i$ correspond to the azimuthal angles. For figure~\ref{fig:multiview} $N=800$ angles are sampled on the unit sphere from the lattice, but not all angles are displayed; Only angles in one hemisphere, which upon folding by mirrors produce images that fall inside the sensor, are retained. Flat circular mirrors with equal radii and corresponding normal angles are appropriately positioned around the object. Note that $N=800$ corresponds to a coarse enough sampling such that none of the mirrors overlap. The sensor dimensions are shown on the focal plane to help visualize how the dense packing of the sensor can work. As seen in figure~\ref{fig:multiview}, the virtual objects are not expected to overlap provided the object dimensions are smaller than the inter-virtual object distance. Moreover it can be seen that, for this view angle sampling and mirror size, no overlap of ray bundles occurs as the light reaches each mirror.

In figure~\ref{fig:multiview} the maximum view angle shown corresponds to $\theta = 55^\circ$, as angles beyond that result in virtual objects that lie outside the field of view of the imaging apparatus. Here the camera has a full frame sensor, with a $f/1.2$ lens of focal length $\unit[62.275]{mm}$ focused for a magnification of $m = 0.215$. The object distance $x$ from the lens is tuned such that the $\theta = 55^\circ$ virtual object lies at the edge of the viewable area. More generally (refer figure~\ref{fig:singleview_folded}),
\begin{align}
\label{eqn:object_distance}
x &= F - \frac{h^*}{\tan{\phi^*}}, \qquad \text{where } \phi^* = \left(\arctan{\frac{h^*}{F}} + \theta^*\right)/2
\end{align}

If $h^*$ corresponds to the maximum height above the optical axis that can be imaged by the lens focused at distance $F$,\footnote{This is distinct from the lens focal length which can be defined as the distance behind the lens where light from infinity is focused at.} then to allow a maximum view angle of $\theta^*$, eq.~\ref{eqn:object_distance} determines the corresponding object distance $x$.

As mentioned in section~\ref{sec:lightfield_single_view}, the defocused image of the object will be formed in the center of the sensor. To avoid overlapping this defocused image with in focus images produced by folded views, no mirrors are placed close to the optical axis, as shown in figure~\ref{fig:multiview}. 

\subsubsection{Related work -- Light field imaging}
\label{sec:lightfield_related_work}

A comprehensive review of light field acquisition and reconstruction techniques can be found in refs.~\cite{8022901} and \cite{Zhou2021}. As presented in ref.~\cite{8022901}, light field acquisition can be split into three main categories: multi-sensor capture, time-sequential capture and multiplexed imaging. Multi-sensor capture involves the use of multiple cameras that each capture one part/direction of the light field. Time-sequential capture involves the use of one sensor that is used to capture different parts of the light field by displacing the sensor and taking images for each direction separately. Finally, multiplexed imaging captures the light field by multiplexing the directions into the spatial or frequency domain. 

The multiple folded view light field imaging concept presented in this text falls into the spatially multiplexed light field imaging category. However, the folding geometry allows a conventional lens to capture light field directions that exceed what is permitted by the lens' aperture. Therefore the angular span captured by this technique exceeds typical spatial multiplexed imaging methods and is in line with the light gathering abilities of multi-sensor imaging devices. Hence the folded view geometry presented here is a hybrid of the two acquisition modalities, and better described as super-aperture multiplexed imaging.

\subsubsection{Implications for cold atom imaging}
\label{sec:lightfield_multi_view_for_atoms}

It is evident from the above discussion that this multiview imaging device is able to capture the light intensity emanating from the object and factorize it into different directions, effectively capturing the light field of the object. This gain in information, along with the gain in light collection, is crucial to the usefulness of the image captured by this device in reconstruction tasks such as those described in section~\ref{sec:results}. This opens up the potential for enhanced light collection, one shot 3D reconstruction and characterization of small scale objects that are dim, fragile, or ephemeral - such as the cold atom clouds seen in atom interferometers. 

The total light collected by the light field imaging device is dictated by the array of mirrors and the lens used to image the plane of virtual objects. In the limit of very small mirrors, the total light collected by the systems tends to 50\%. In this regime each view has a severely reduced field of view and collects a small fraction of the total light. Note that the maximum is 50\% since the mirrors are arranged to cover one hemisphere around the object.

In the limit of large mirrors, each view collects as much light as the lens' aperture maximally allows. Assuming the imaging apparatus is tuned for atom cloud imaging as described in section~\ref{sec:conventional_imager}, the light field imager can collect approximately $m$ times the light that the conventional imager collects, where $m$ is the number of views folded with the mirrors. This offers a potential order of magnitude gain in the light collected depending on the size and number of mirrors used. This is done while simultaneously ensuring that the depth of field of the system is acceptable since the same parameters for the imaging apparatus are used as discussed in section~\ref{sec:conventional_imager}; Each view has an effective F-number greater than or equal to the single-view conventional setup and thus necessarily has a greater or equal depth of field. An experimental realization of this idea is presented below in section~\ref{sec:experiment} where $n=111$.
\section{Experiment}
\label{sec:experiment}


\subsection{Demonstrator Design}
\label{sec:demo_design}



Using 3D-printing technology, we have built a proof-of-concept demonstrator of the imaging system described in section~\ref{sec:optical}, to demonstrate the multi-view imaging and the 3D-reconstruction capabilities. The success of this demonstrator also implies that the system can be deployed quickly and cost-effectively in other experiments. We discuss the details of the demonstrator design parameters in this section.

In realistic experiments with cold atom clouds, the environment around the nominal atom cloud position is often populated with various laser beams for magneto-optical traps and other purposes. Also, the camera and the lens would typically lie outside of the vacuum chamber that the atom clouds are located in. Therefore, based on estimated specifications for MAGIS-100, we require the closest mirrors and the lens, respectively, to be at least $\unit[3]{cm}$ and $\unit[15]{cm}$ away (along the optical axis) from the nominal object position. This constraint sets the maximum view angle to be $\theta = 55^\circ$ for a $\unit[35]{mm}$ full-frame sensor.

For the demonstrator, we use $\unit[5]{mm}$ mirrors, since these are the smallest mirrors available off-the-shelf. This mirror diameter is large enough to allow for a sufficient field-of-view for an $\mathcal{O}(\unit[1]{mm})$ target object, while small enough to pack the range of solid angle efficiently and accommodate up to $n = 111$ views. (A small central region is not filled with the mirrors, because the defocused object will interfere with them.) The results in this paper are produced with a demonstrator holding $n = 90$ mirrors.

For the camera, we choose the QHY600 camera (typically used for astrophotography), as it provides a full-frame, back-illuminated sensor with a small pixel size ($\unit[3.76]{\mu m}$) and 16-bit analog-to-digital bit depth. For the lens, we prioritize having a wide range of $f$-number to operate in, while targeting reasonable magnification and object distance constraints, explained in section~\ref{sec:optical}. Among the commercially available lens options respecting these constraints, we choose the Nikkor Z $\unit[58]{mm}$ $f/0.95$ S lens. This is one of the largest aperture lenses available, with a good, well-documented off-axis performance. 

\begin{figure}[t]
    \centering
    \begin{subfigure}[t]{0.63\textwidth}
        \centering
        \includegraphics[height=2.5in]{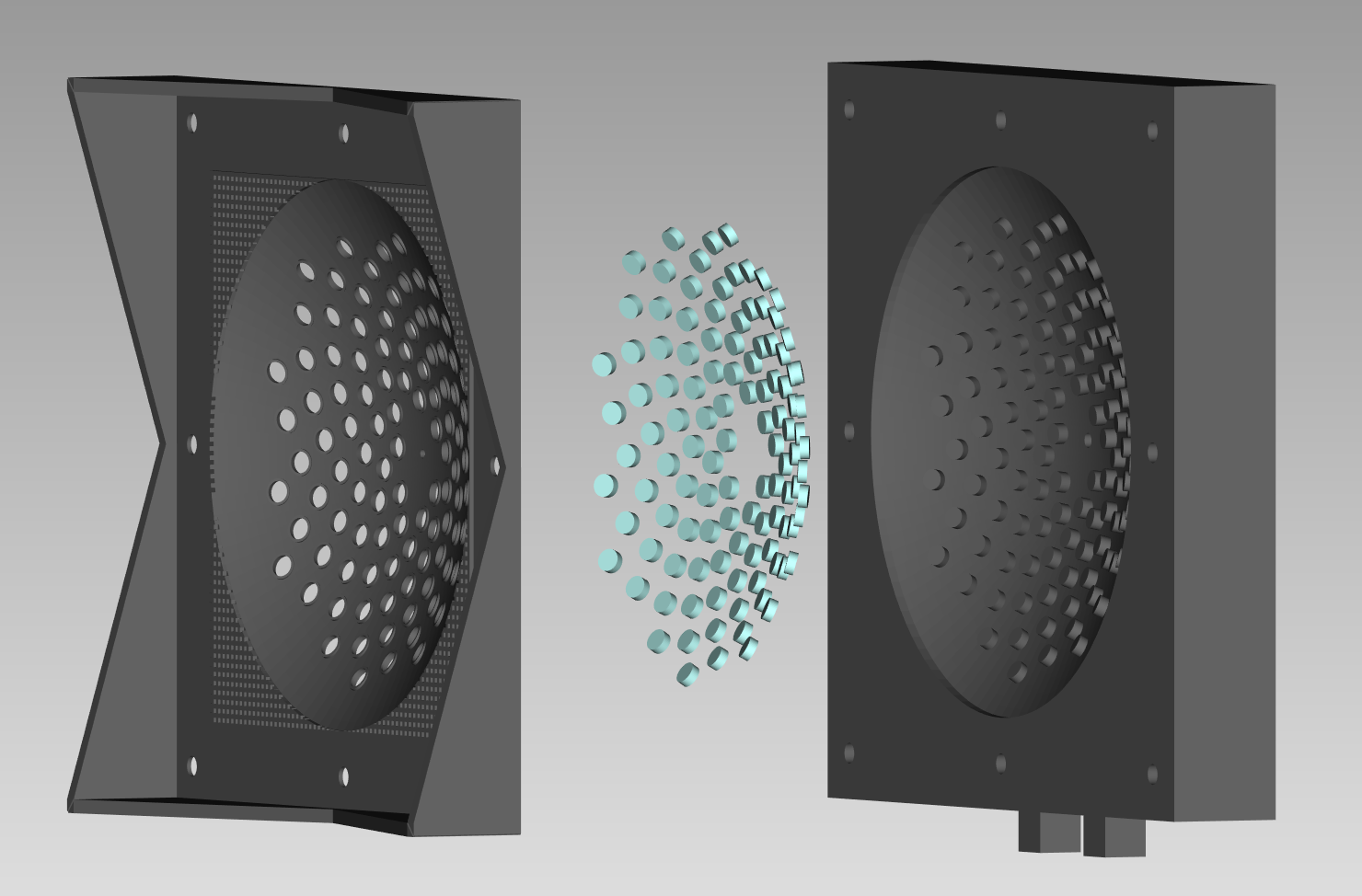}
        \caption{Exploded view}
        \label{fig:demonstrator_cad_exploded_view}
    \end{subfigure}
    \begin{subfigure}[t]{0.33\textwidth}
        \centering
        \includegraphics[height=2.5in]{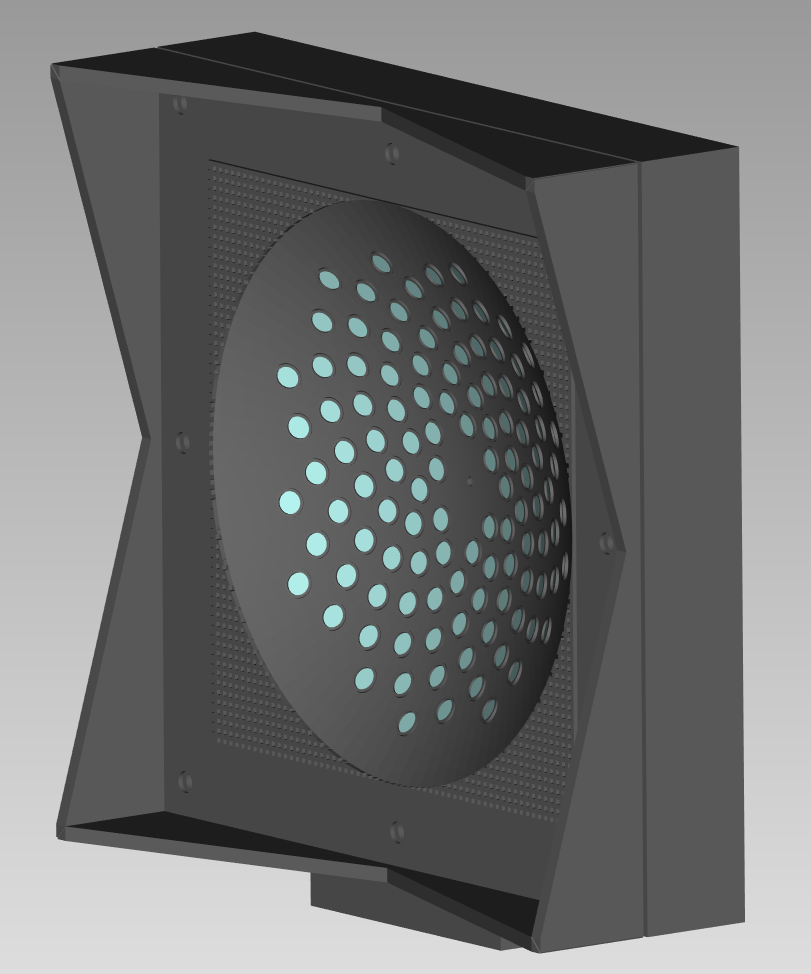}
        \caption{Assembled view}
        \label{fig:demonstrator_cad_assembled_view}
    \end{subfigure}
    \\
    \centering
    \begin{subfigure}[t]{0.48\textwidth}
        \centering
        \includegraphics[height=1.8in]{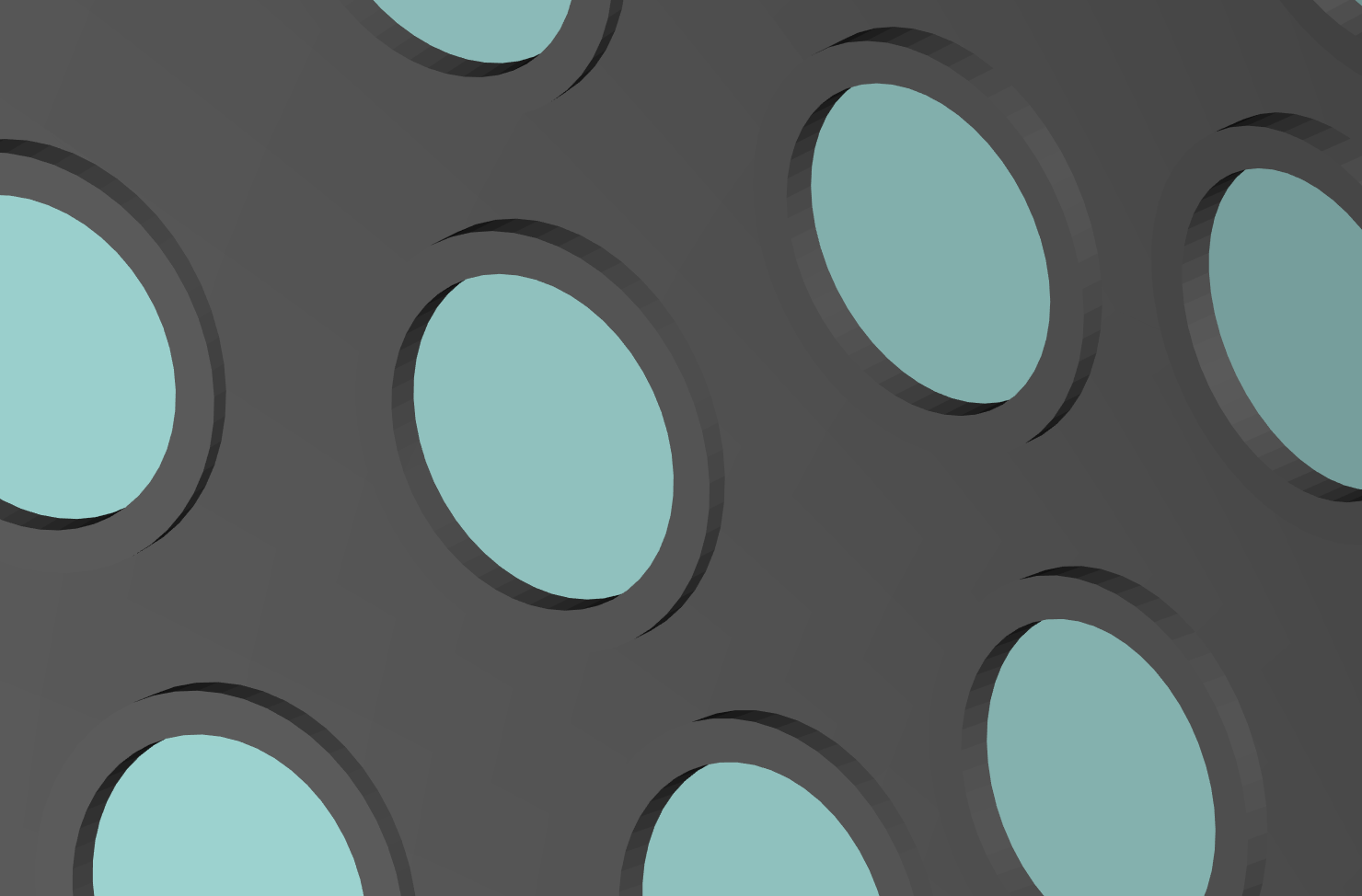}
        \caption{Front board and mirrors}
        \label{fig:demonstrator_cad_zoom_front_board}
    \end{subfigure}
    \begin{subfigure}[t]{0.48\textwidth}
        \centering
        \includegraphics[height=1.8in]{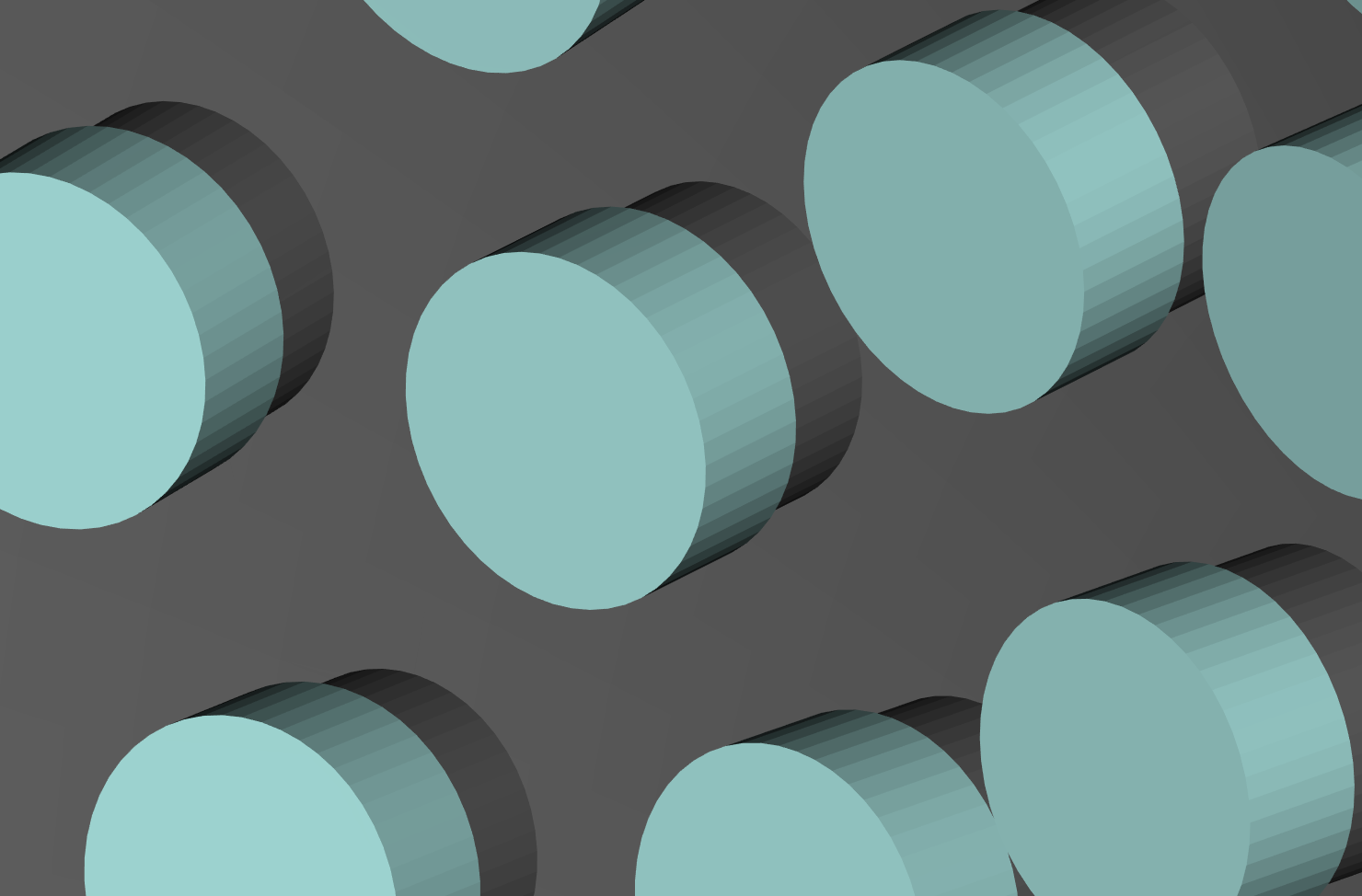}
        \caption{Back board and mirrors}
        \label{fig:demonstrator_cad_zoom_back_board}
    \end{subfigure}
    \caption{Various views of the demonstrator CAD model. The teal disks are the mirrors. (a) The black parts to the left and the right of the mirrors are the ``front'' board and the ``back'' board, respectively. (c) A zoomed-in view, showing each mirror being pushed against a ring-like structure on the front board. (d) A zoomed-in view without the front board, showing each mirror being pushed by a cylindrical structure on the back board.}
    \label{fig:demonstrator_cad}
\end{figure}

\begin{figure}[t]
    \centering
    \includegraphics[width=0.5\textwidth]{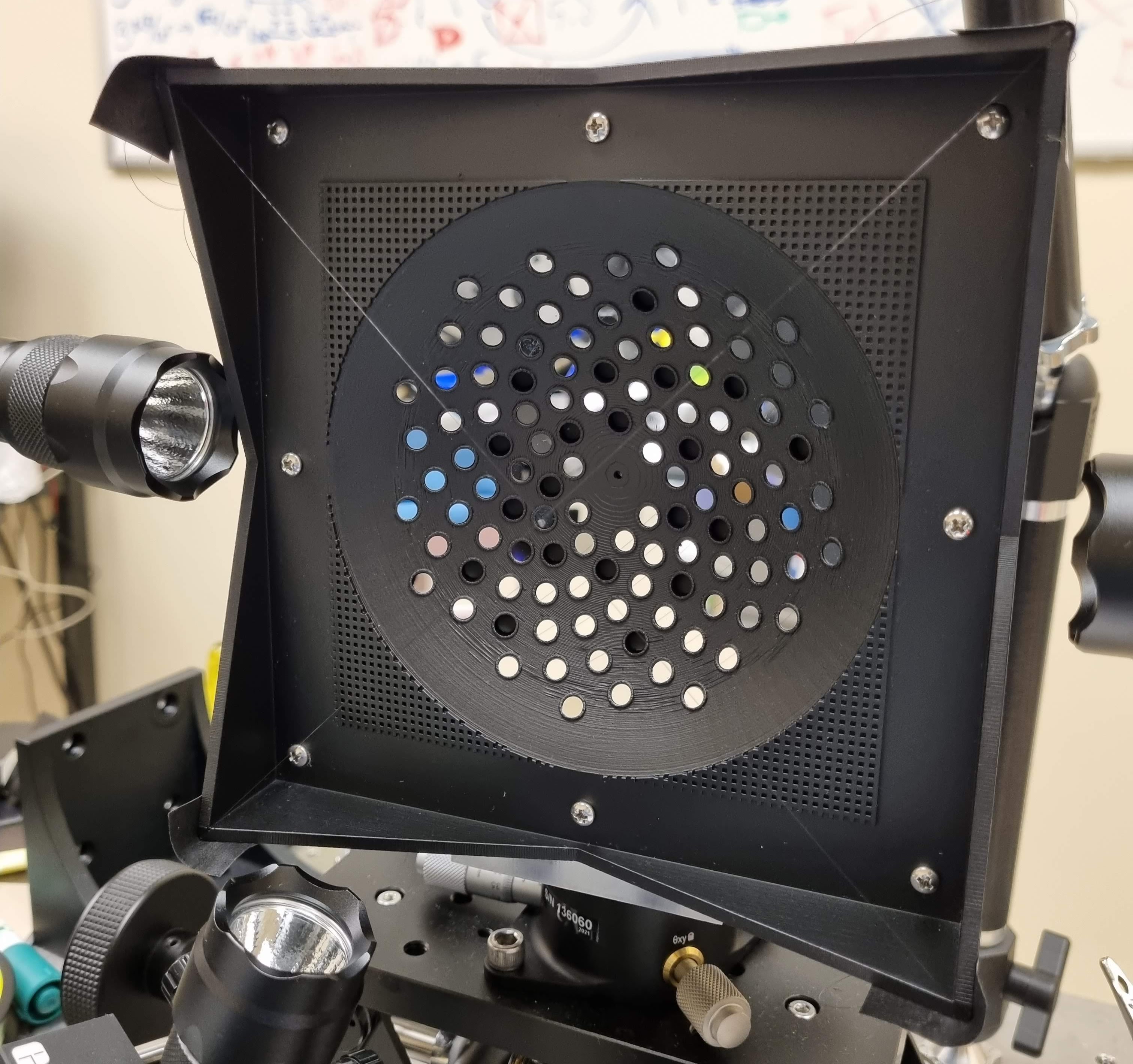}
    \caption{Photo of the demonstrator. A pair of thin (0.005") fibers are used to create a crosshair that defines the nominal target object position and hold the object in place.}
    \label{fig:demonstrator_lab_photo}
\end{figure}

The array of mirrors are held in their respective positions and angles by two 3D-printed boards as shown in figure~\ref{fig:demonstrator_cad_exploded_view}. The two boards and the mirrors are tightened using bolts and nuts along the edges of the demonstrator. The ``front'' board has an array of small rings that define the mirror positions and angles, while the ``back'' board has an array of cylinders that push and apply pressure to the mirrors, so that the mirrors are not floating freely. See figures.~\ref{fig:demonstrator_cad_zoom_front_board} and \ref{fig:demonstrator_cad_zoom_back_board}. The nominal thickness of the front board is $\unit[1]{mm}$ which is thinner than the nominal thickness of the mirrors $\unit[1.5]{mm}$, ensuring that the cylindrical knobs from the back board are pushing the mirrors. The front board has angle brackets around its edges, and their dimensions are set such that the front-most plane defines the nominal imaging target position along the optical axis ($x = 0$). The corners of the angle brackets are engraved with grooves, and thin (0.005") fibers can be used to place the target object at the precise nominal position using these grooves. See figure~\ref{fig:demonstrator_lab_photo}.

 The demonstrator is 3D-printed using stereolithography (SL) technology with 0.004" $Z$-resolution and 0.010" $XY$-resolution, and the optical axis of the demonstrator is aligned as the $Z$-axis during 3D printing. This sets the positional and the angular resolutions of the mirror holes on the front board. In particular, these resolutions are worse at larger view angles. At the most extreme view angle $\theta = 55^\circ$, the mirror surface normal is at $35^\circ$ from the optical axis ($Z$-axis in manufacturing), and this normal surface has a combined resolution of $\unit[160]{nm}$. For $\unit[5]{mm}$ diameter mirrors, this corresponds to an angular resolution of $1.8^\circ$. Image-level comparisons with our simulation confirms that the errors are smaller than these expectations.

\begin{figure}
    \centering
    \includegraphics[width=0.5\textwidth]{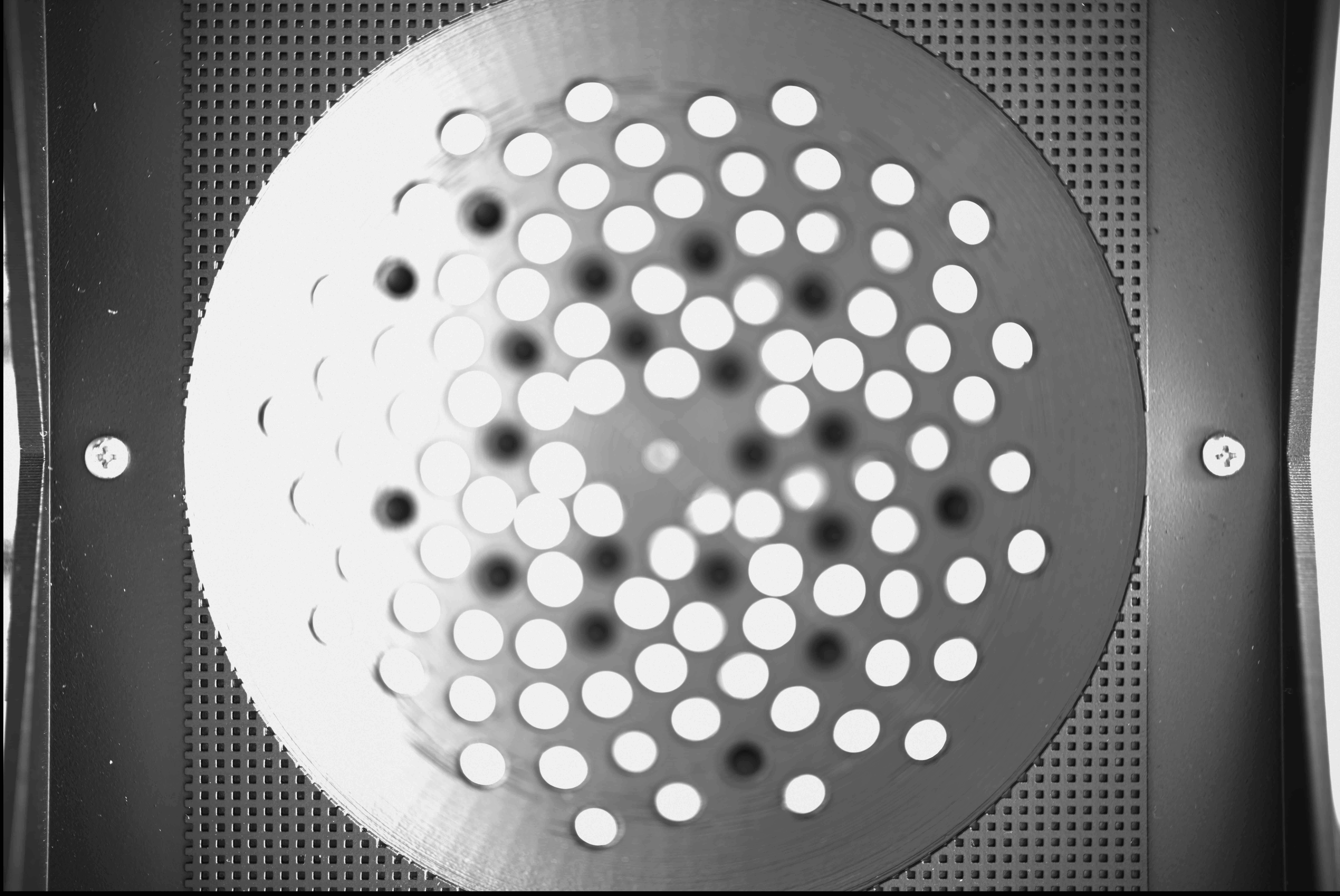}
    \caption{Image of the alignment grid of the demonstrator. The entire grid is focused within DoF of $\unit[2]{mm}$.}
    \label{fig:demonstrator_grid_alignment}
\end{figure}

The front board also has an alignment grid, a $\unit[12]{cm} \times \unit[12]{cm}$ square grid of $\unit[1]{mm}$ wide and $\unit[1]{mm}$ deep lines printed on its flat surface. We align the overall demonstrator perpendicular to the optical axis by ensuring that the entire alignment grid is simultaneously in focus while operating at low depth-of-field (DoF). Operating our camera and lens at $f / 1.1$ and experimenting with the focus of the grid pattern, we find the effective DoF to be approximately $\unit[2]{mm}$, which means that this method will align the demonstrator within $\pm \unit[1/60]{rad}$ ($\pm 0.95^\circ$) from the exact perpendicular plane. See figure~\ref{fig:demonstrator_grid_alignment} for an example image of the front board alignment grid.

\subsection{Imaging with Demonstrator}
\label{sec:imaging}

To test our imaging system, we use the demonstrator to take images and perform 3D reconstructions of a test object. The test object is a cubic shape that is $\mathcal{O}(\unit[1]{mm})$ in the overall sizes and $\mathcal{O}(\unit[100]{{\mu}m})$ in feature sizes. The test object is 3D-printed with projection micro-stereolithography (P{\textmu}SL) technology, which enables precise manufacturing of micron-scale features.

The test object is manufactured using HTL resin \cite{BMF_HTL_resin}. Under $\unit[405]{nm}$ light, this material fluoresces broadly over $\unit[430]{nm}$ -- $\unit[550]{nm}$. Therefore, we capture images of the test object illuminated only using $\unit[405]{nm}$ LED lights, with a $\unit[450]{nm}$ filter placed in front of the camera sensor. The object is illuminated from five different directions: top, bottom, left, right, and behind through a small central hole in the demonstrator. While this set-up does not illuminate the object uniformly and the object does not irradiate uniformly over $4\pi$ solid angle, our reconstruction method is able to capture view-angle dependent light field.

\section{Simulation and reconstruction}
\label{sec:sim_reco}


For analysis and reconstruction using the system described in sections~\ref{sec:optical} and \ref{sec:experiment}, we implement the same design in a custom simulation framework. This framework is based on geometric optics, which automatically simulates effects such as depth of field, and can be thought of as a mapping $S_{\theta} \colon \rho(x) \mapsto y$, where $\theta$ is a set of parameters of the imaging system, e.g. the mirror positions/angles and the lens focal length. The objects to be imaged, $\rho(x)$, are densities dependent on 3D spatial coordinates $x \in \mathbb{R}^3$, which may be represented by analytical functions, voxels, meshes, or neural networks, and the outputs, $y \in \mathbb{R}^2$, are images captured by a user-defined optical setup. We have built this framework to be fully differentiable, allowing for gradient-based reconstruction of the simulation inputs (via gradients with respect to parameters of $\rho(x)$) and gradient-based calibration (via gradients with respect to $\theta$). A demonstration of these latter properties is the subject of future work. A review of physically-based rendering frameworks, which have informed our design, can be found in ref.~\cite{pbrt2016}.

For the system presented here, reconstruction is done on the level of the views described in Section~\ref{sec:optical}. In terms of notation, we define a view $v_{i}$ as the object image corresponding to a mirror $m_{i}$. In practice a $H \times W$ pixel patch from the full image captured by the camera, where $H$ and $W$ are respectively the height and width of the patch in pixels. For a given image $y$, we extract a set of $n$ such folded views with a patch finding algorithm, yielding a dataset $\mathcal{D} = \{v_i\}_{i=1}^n$, where each view is a set of of pixels $j,k$ together with their measured intensity $I_{i, j}$, that is $v_{i} = \{j_i, k_i, I_{j_i, k_i}\}_{j, k = 1}^{H \times W}$. Such views may be matched to the corresponding views in simulation, and the aim of reconstruction is to learn a 3D input model $\rho(x)$ for the simulator such that the corresponding simulation output $y$ closely matches the captured image. Specifically, for simulated intensities $S_{\theta}(\rho(x))_{j, k}$ at pixels $j,k$ and target intensities $I_{j, k}$, we optimize $\arg \min_{\rho(x)} \sum_i \sum_{j_i} \sum_{k_i}$ $||S_{\theta}(\rho(x))_{j_i, k_i} - I_{j_i, k_i}||_2^2$ with gradient descent, where the sums are over views and pixels in each view. 

As will be shown in section~\ref{sec:results}, we consider the reconstruction of two categories of objects: simulated atom clouds, motivating future applications of our sensing device, and 3D printed fluorescent objects, used to assess the capability of the demonstrator described in section~\ref{sec:demo_design}. For both applications, we represent the objects (i.e. their densities $\rho(x)$) using neural networks. We benefit from developments in the related applications of neural scene representation~\cite{SRN,NeRF} and, more broadly, implicit neural representation~\cite{siren, neuralFields}. Implicit neural representations have seen some exploration for scientific imaging, including for CT reconstruction~\cite{9606601,9709943} and for synthetic aperture sonar reconstruction~\cite{9705799}. 

The atom cloud reconstruction is a tomographic reconstruction. We model the volume with a neural network that takes a position as input and outputs a density at that point in space, similar to recent methods in volumetric rendering~\cite{pbrt2016,NeRF, autoint2021}. The weights of the neural network are optimized with gradient descent with the procedure described above so that after training, the neural network models the density of the target volume. For this work, we use SIREN~\cite{siren}, a recently proposed architecture for modeling 3D densities, which uses neural networks with sinusoidal activation functions to learn high quality volumetric signals.

In contrast to the clouds, which have no well defined surfaces, the semi-opaque 3D printed objects used for the demonstrator have sharply defined surfaces. We therefore use a method predominantly focused on surface reconstruction, NeuS~\cite{NeuS}. This method represents a surface via two neural networks: a signed distance function (SDF) and a view dependent color. The learned SDF is used to construct a density highly peaked about its 0 value (the learned surface), and resulting views are constructed via standard volumetric rendering with this highly peaked density. The neural network predicting view dependent color is needed due to the non-uniform illumination of the object. Note that the optics simulation of the physical demonstrator is used for the generation of views at each training stage. While simulation parameters are kept fixed to nominal values during training, co-optimization of parameters and model is an interesting direction for future work.
\section{Results}
\label{sec:results}

\subsection{Simulation results}

\begin{figure}[ht]
    \centering
    \includegraphics[height=5in]{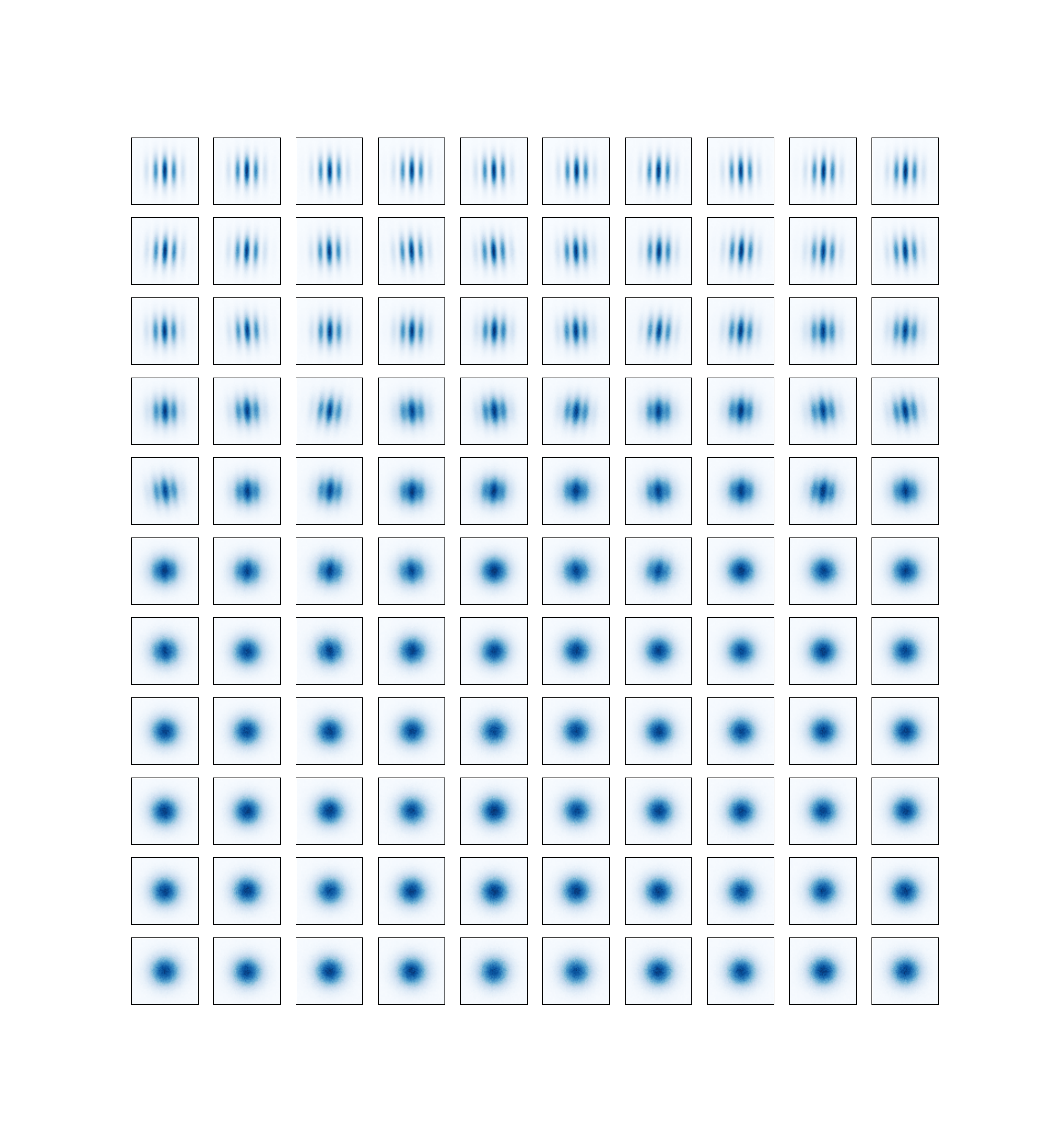}
    \caption{Grid of views extracted from a simulated light field 60MP image of an atom cloud imaged over $\unit[10]{\mu s}$.}
    \label{fig:simulated_image_cloud}
\end{figure}

To demonstrate the tomographic capabilities of our light field device, we simulate the imaging of an atom cloud with a sine-wave density modulation fringe pattern. The number of atoms is sampled from a Poisson distribution with $\lambda=10^6$ atoms, where each atom emits $10^8$ photons per second. Figure~\ref{fig:simulated_image_cloud} shows 110 (out of 111) views extracted from a simulated light field image, using an exposure time of $\unit[10]{\mu s}$.

\begin{figure}[ht]
    \centering
    \includegraphics[height=5in]{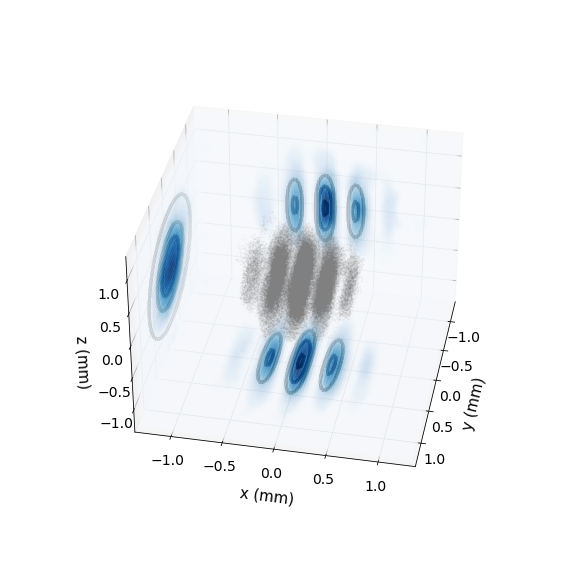}
    \caption{Data sampled from the learned 3D cloud model in gray, and the learned 2d marginals in blue against the target density in black (the 40, 68 and 95\% contours are shown).}
    \label{fig:learned_cloud}
\end{figure}

For reconstruction, we model the atom cloud density with a neural network and use the views from figure~\ref{fig:simulated_image_cloud} in order to train it as described in section~\ref{sec:sim_reco}. Figure~\ref{fig:learned_cloud} shows the learned density after training the neural network for 5,000 epochs. It can be observed that the input views acquired in a single shot allow to faithfully reconstruct the atom cloud without using domain knowledge or regularization.

\subsection{Captured data}


\begin{figure}[h]
    \centering
    \includegraphics[height=5in]{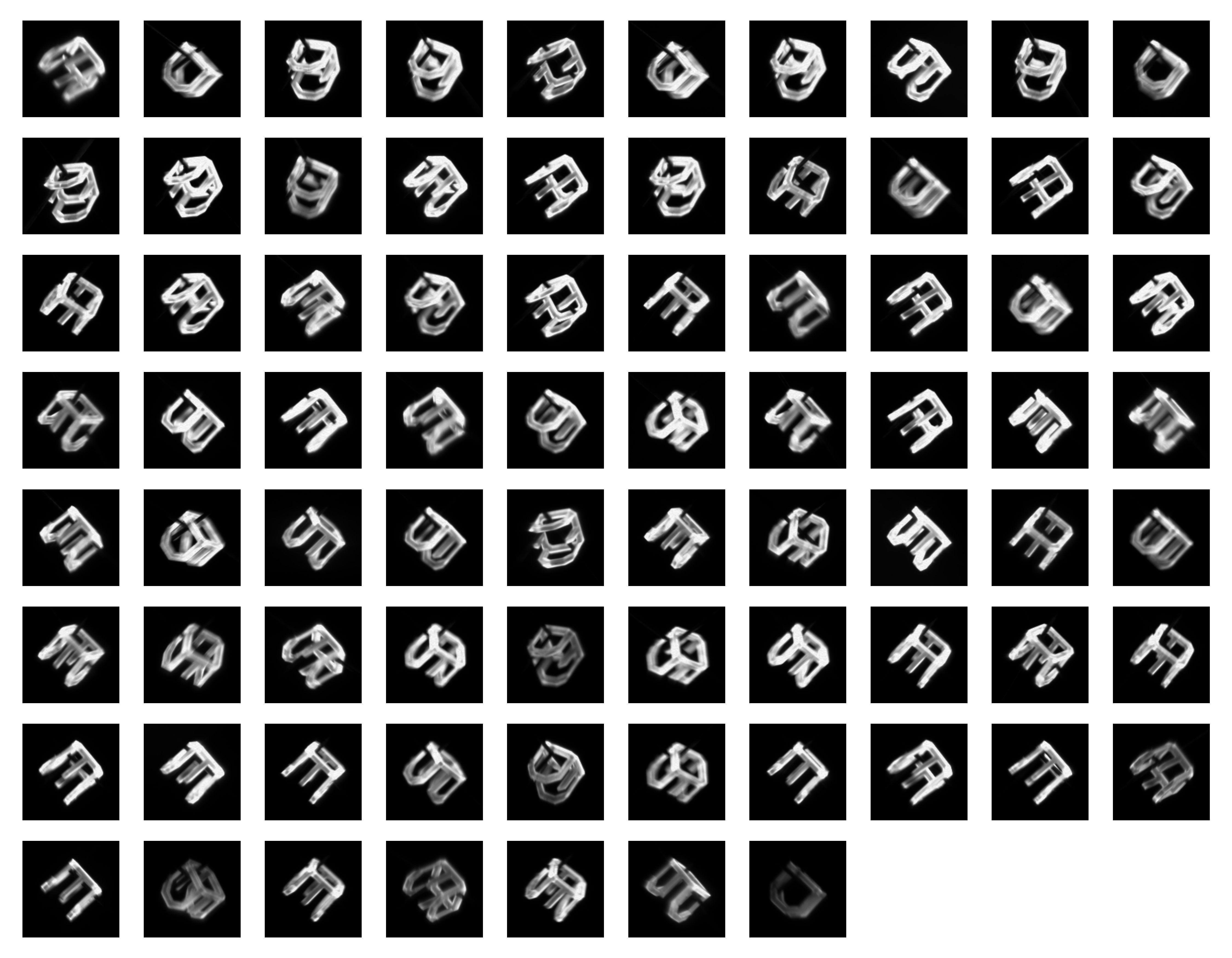}
    \caption{Grid of training views extracted from a 60MP image of the test object, taken in a single shot.}
    \label{fig:real_img_grid}
\end{figure}

Images captured by the demonstrator are $9600 \times 6422$ pixels, with an embedded set of views of the imaged object (nominally $n$ views for $n$ mirrors in the device). As mirrors within the constructed imaging apparatus may differ slightly from the expected position due to imperfect construction or alignment, mirrors in simulation and observed data must be matched before per-view comparisons can be performed. We run a patch finding algorithm based on the contour and rectangle finding tools in OpenCV~\cite{openCV} to extract $H\times W$ pixel patches corresponding to each view. Expected view positions for each mirror are found in simulation by tracing a ray from the expected object position to each mirror center, and following this ray to its intersection with the sensor. These intersection points are compared with the view centroids found by the patch finding, and the centroid closest to the ray intersection point from a given mirror is identified as the view corresponding to that mirror. The set of views used for the reconstruction results presented here are shown in figure~\ref{fig:real_img_grid}. Out of the 90 views instrumented in the demonstrator, 86 views are found by the patch finding, with the others falling below a small intensity threshold. To ensure a good match with simulation, a threshold of 350 pixels is set on the distance between patch centroid and the corresponding mirror position matched as described above. With this threshold, 82 views are available for reconstruction. While running a reconstruction is feasible with these 82 views, quality is improved by discarding an additional 5 views which are poorly centered (from a shifting in the bounding box due to illumination of the suspending thread) or poorly illuminated. The issue of thread illumination is particular to our demonstration setup, and the impacted views are recoverable with further post-processing. However, the inclusion of these recovered views did not improve reconstruction quality significantly. The views which are poorly illuminated or are not found at all by the patch finding are likely the result of individual mirror misalignments, and may be improved with a more sophisticated mirror mounting or higher precision construction of the device.

\subsection{Experimental results}
\label{results_exp_results}

\begin{figure}[h]
    \centering
    \includegraphics[width=\textwidth]{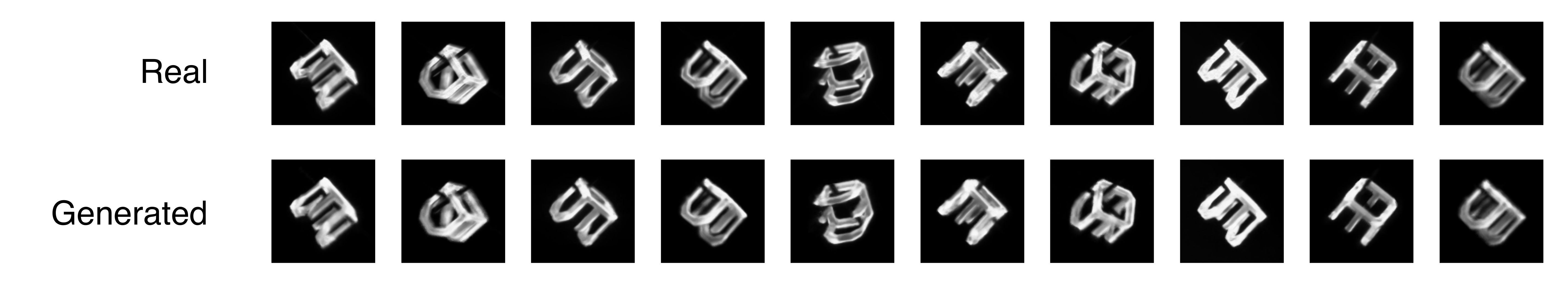}
    \caption{Comparison of subset of training (real) views to the corresponding subset of generated views from the learned NeuS reconstruction. Results are visually nearly identical between real captured images and their generated counterparts, demonstrating reasonable closure of the model.}
    \label{fig:real_gen_compare}
\end{figure}

For the purposes of reconstruction, the patch in data is assumed to be at the corresponding nominal position from simulation; namely the set of $H \times W$ pixels centered at the expected view position described above. These patches, with the corresponding identified mirrors and pixel positions, are used as the inputs for the reconstruction methods described in section ~\ref{sec:sim_reco}. For the results presented here, patches of $130 \times 130$ pixels are used. The NeuS reconstruction is trained for 100000 iterations on the 77 views shown in figure ~\ref{fig:real_img_grid}, where a training iteration corresponds to a sampling of 512 pixels from a view, within a sweep across randomized views. No explicit masking, i.e. supervised information about the object location in each view, is used and the distinction between foreground and background is learned. Following the NeuS implementation, the background is modeled by NerF++~\cite{Nerfpp}. Full hyperparameter details are in appendix~\ref{appendix:neus_params}. Network and training settings closely follow those used in ref.~\cite{NeuS}. Figure~\ref{fig:real_gen_compare} compares a subset of real views with the corresponding views generated from the learned NeuS model. Results are visually nearly identical between the real and generated images, demonstrating the good convergence of this trained model. The full grid of generated training views is available in appendix~\ref{appendix:image_grids}.

\begin{figure}[h]
    \centering
    \begin{subfigure}[t]{0.32\textwidth}
        \includegraphics[width=\textwidth]{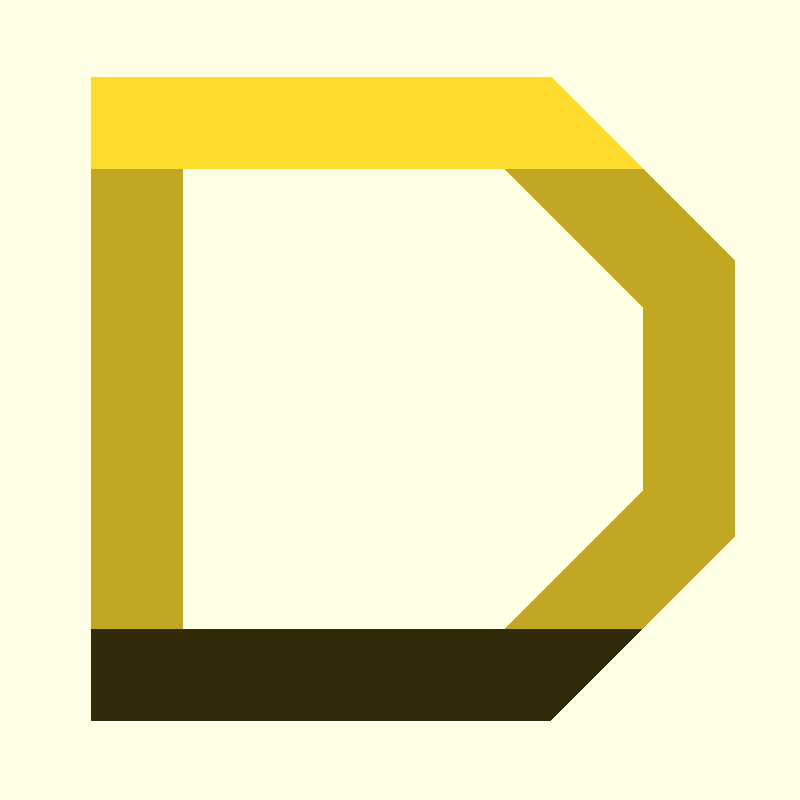}
    \end{subfigure}
    \begin{subfigure}[t]{0.32\textwidth}
        \includegraphics[width=\textwidth]{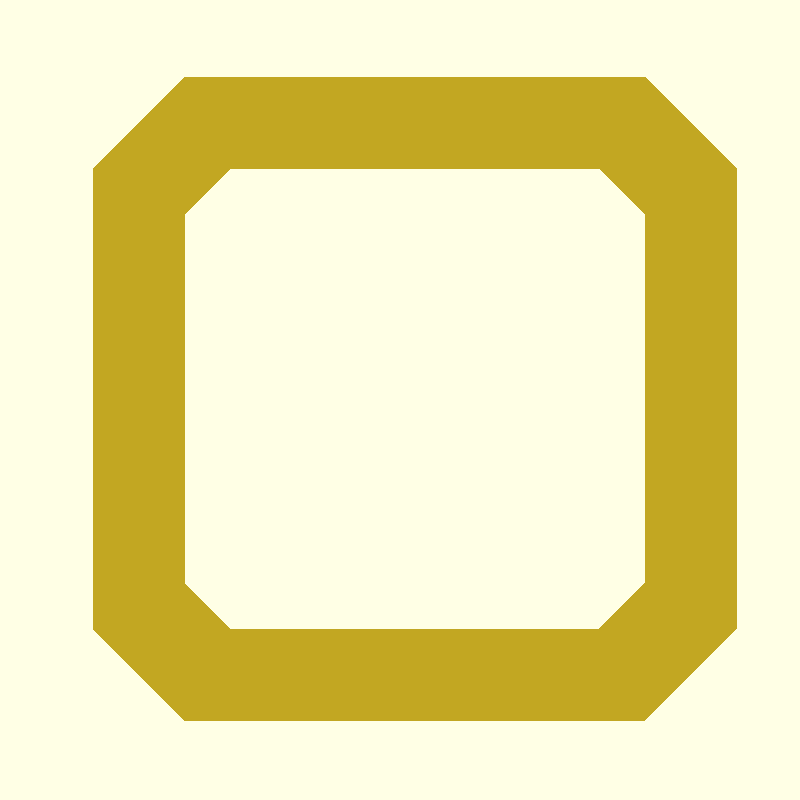}
    \end{subfigure}
    \begin{subfigure}[t]{0.32\textwidth}
        \includegraphics[width=\textwidth]{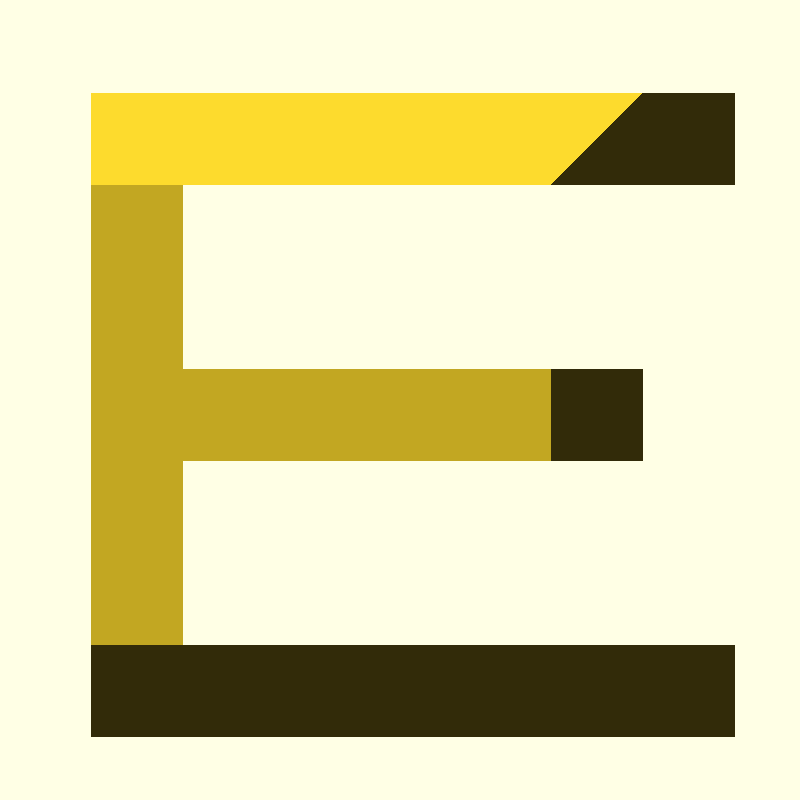}
    \end{subfigure}
    \\
    \begin{subfigure}[t]{0.32\textwidth}
        \includegraphics[width=\textwidth]{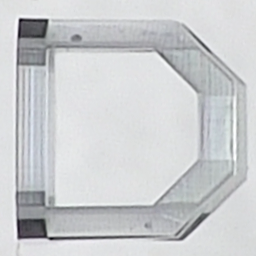}
    \end{subfigure}
    \begin{subfigure}[t]{0.32\textwidth}
        \includegraphics[width=\textwidth]{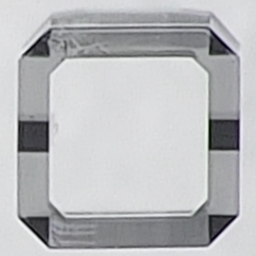}
    \end{subfigure}
    \begin{subfigure}[t]{0.32\textwidth}
        \includegraphics[width=\textwidth]{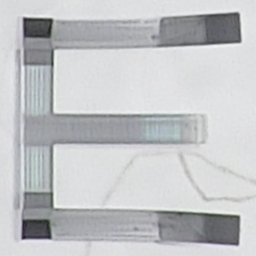}
    \end{subfigure}
    \\
    \begin{subfigure}[t]{0.32\textwidth}
        \includegraphics[width=\textwidth]{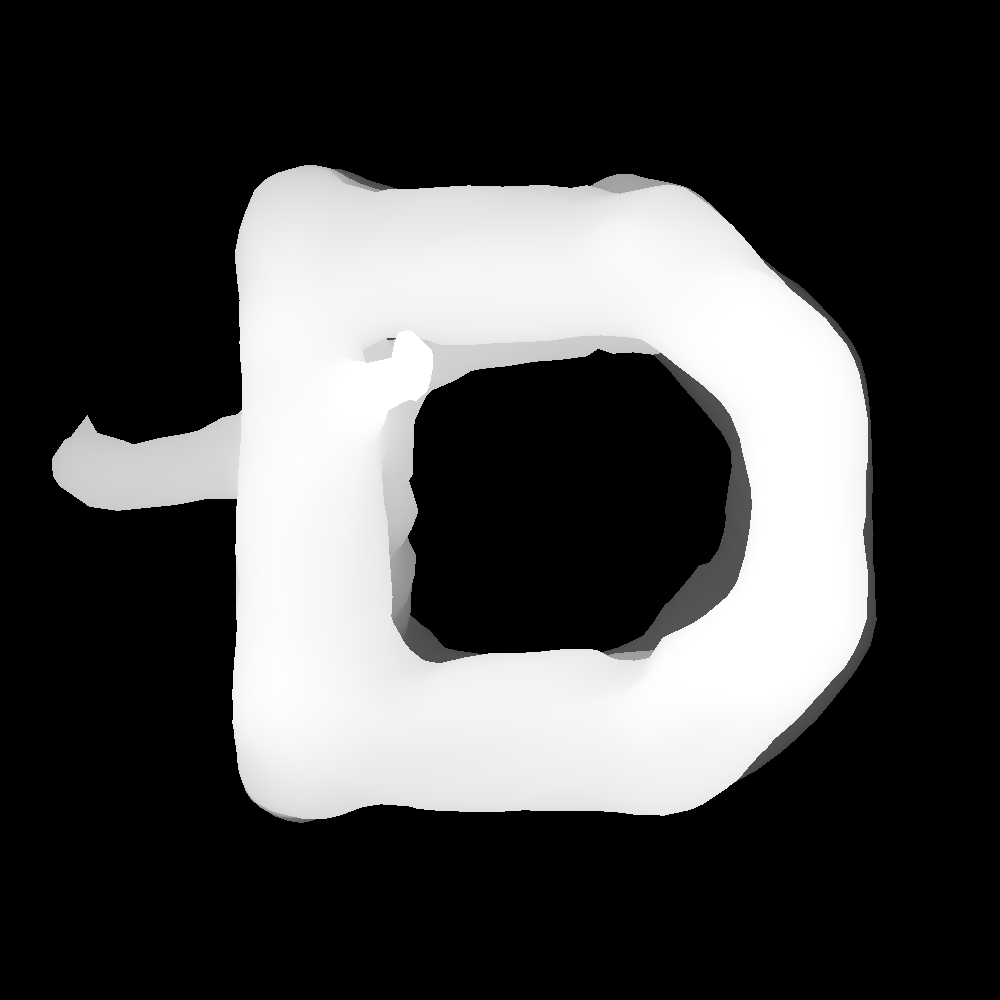}
    \end{subfigure}
    \begin{subfigure}[t]{0.32\textwidth}
        \includegraphics[width=\textwidth]{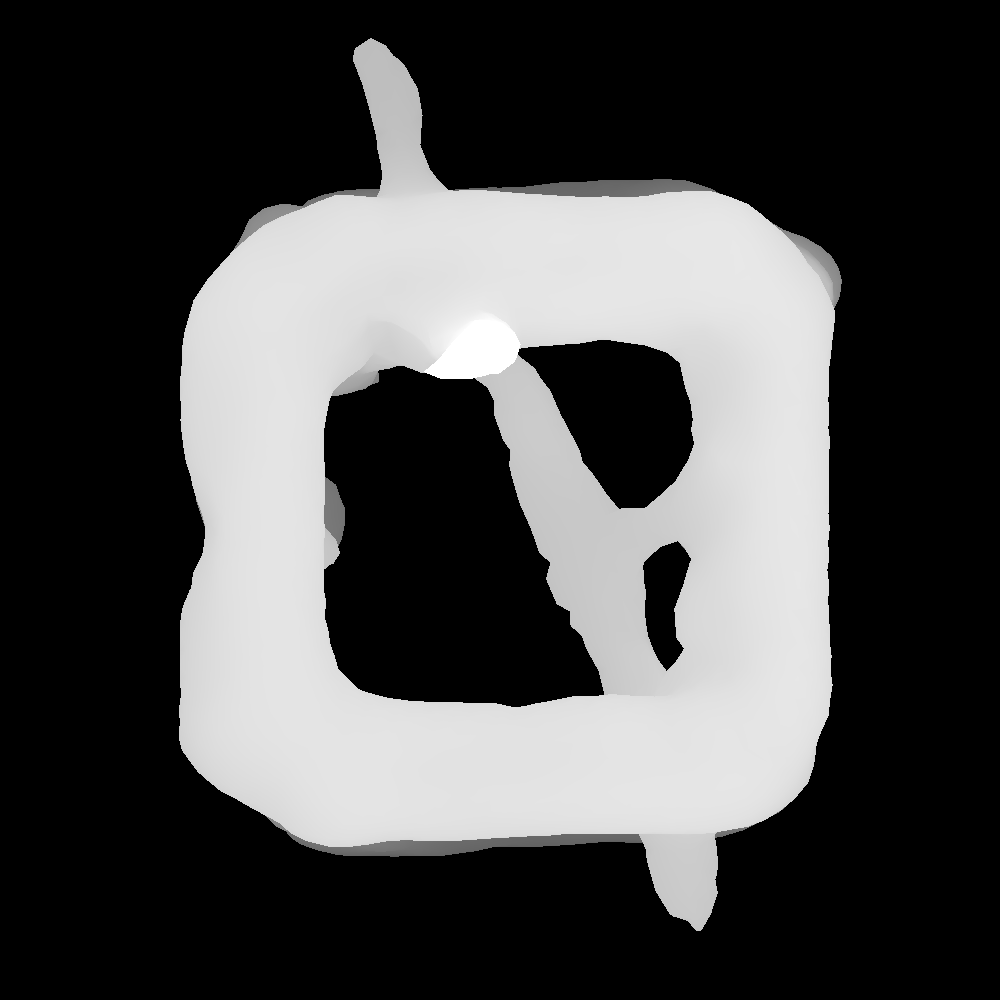}
    \end{subfigure}
    \begin{subfigure}[t]{0.32\textwidth}
        \includegraphics[width=\textwidth]{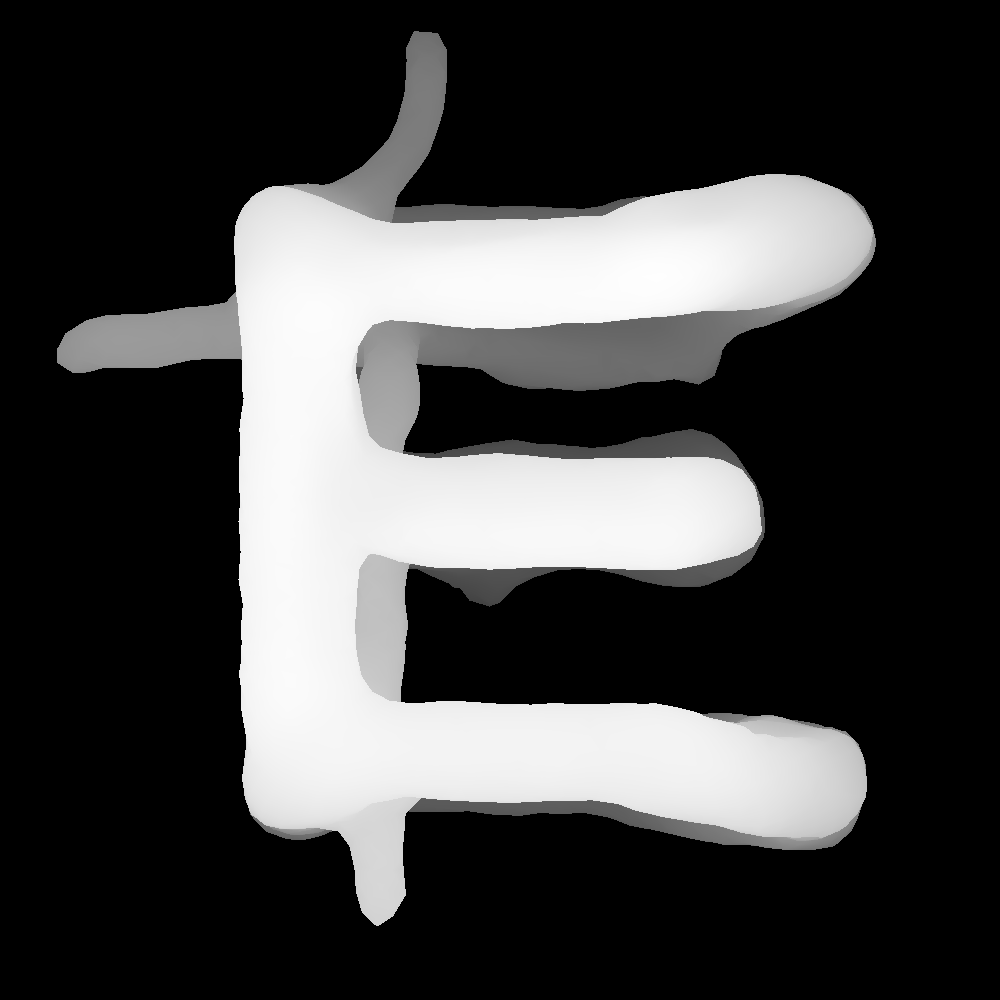}
    \end{subfigure}
    \caption{Reconstruction of the ``DOE cube'' test object. The envelope dimension of the object is $(\unit[1]{mm})^3$, and the letter is $\unit[142]{\mu m}$ thick. Top: CAD design, middle: microscope images of a 3D printed object, bottom: depth map of surface reconstructed using 77 views from our system. Different columns correspond to different views of the same objects. Note that the reconstructed results contain artifacts from the $\unit[120]{\mu m}$ thread used to suspend the object, as well as from the limited range of viewing angles.}
    \label{fig:DOE_cube_results}
\end{figure}

\begin{figure}[h]
    \centering
    \begin{subfigure}[t]{0.32\textwidth}
        \includegraphics[width=\textwidth]{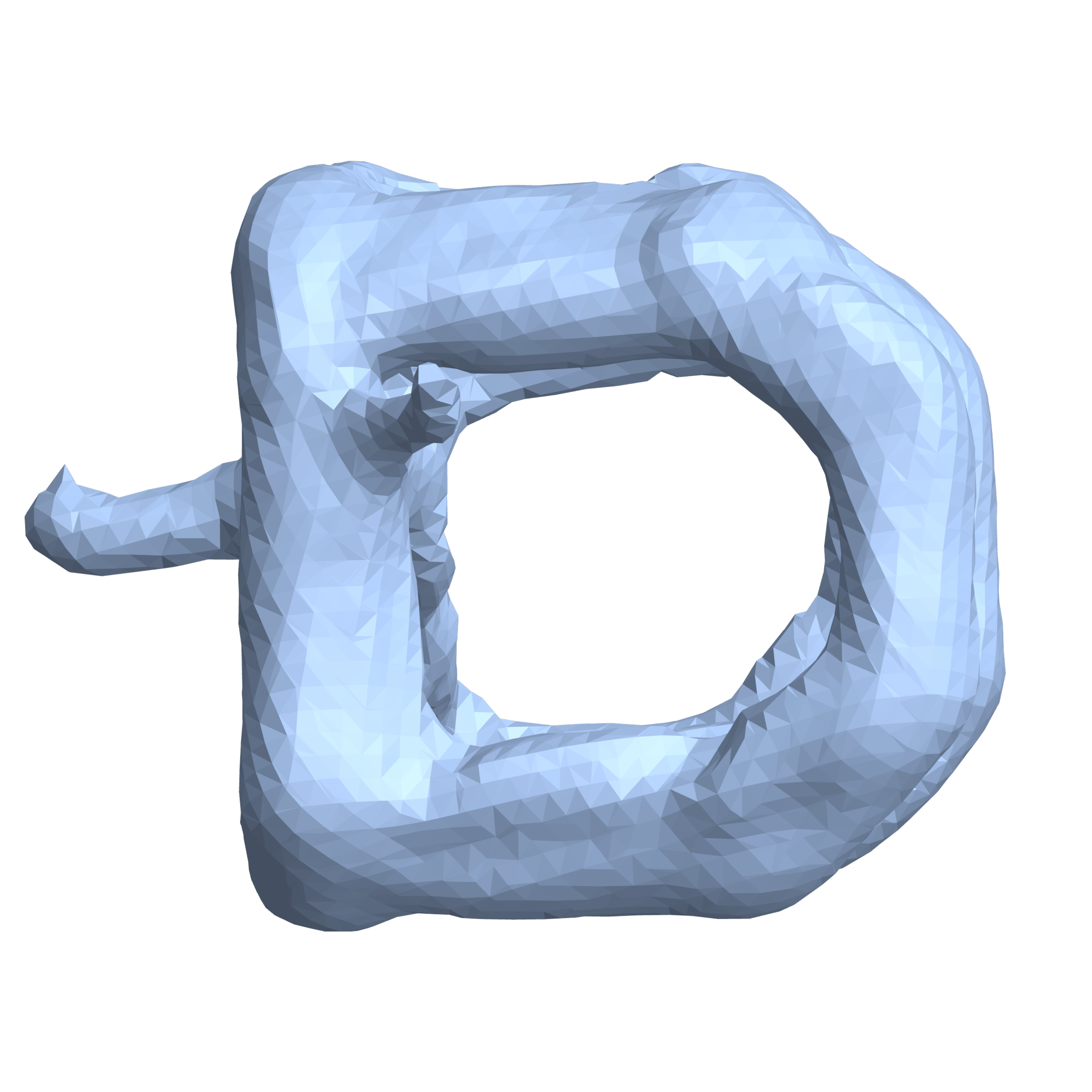}
    \end{subfigure}
    \begin{subfigure}[t]{0.32\textwidth}
        \includegraphics[width=\textwidth]{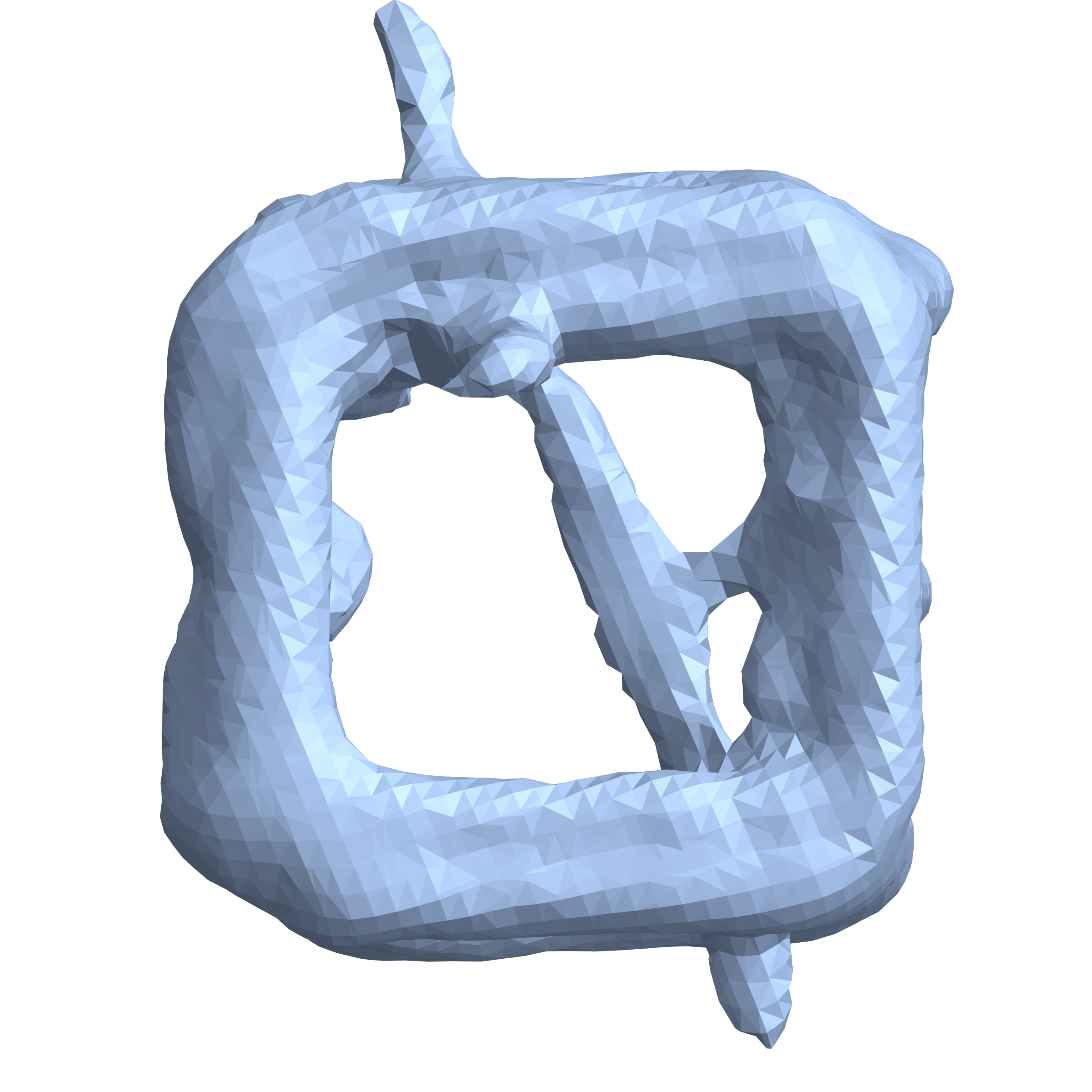}
    \end{subfigure}
    \begin{subfigure}[t]{0.32\textwidth}
        \includegraphics[width=\textwidth]{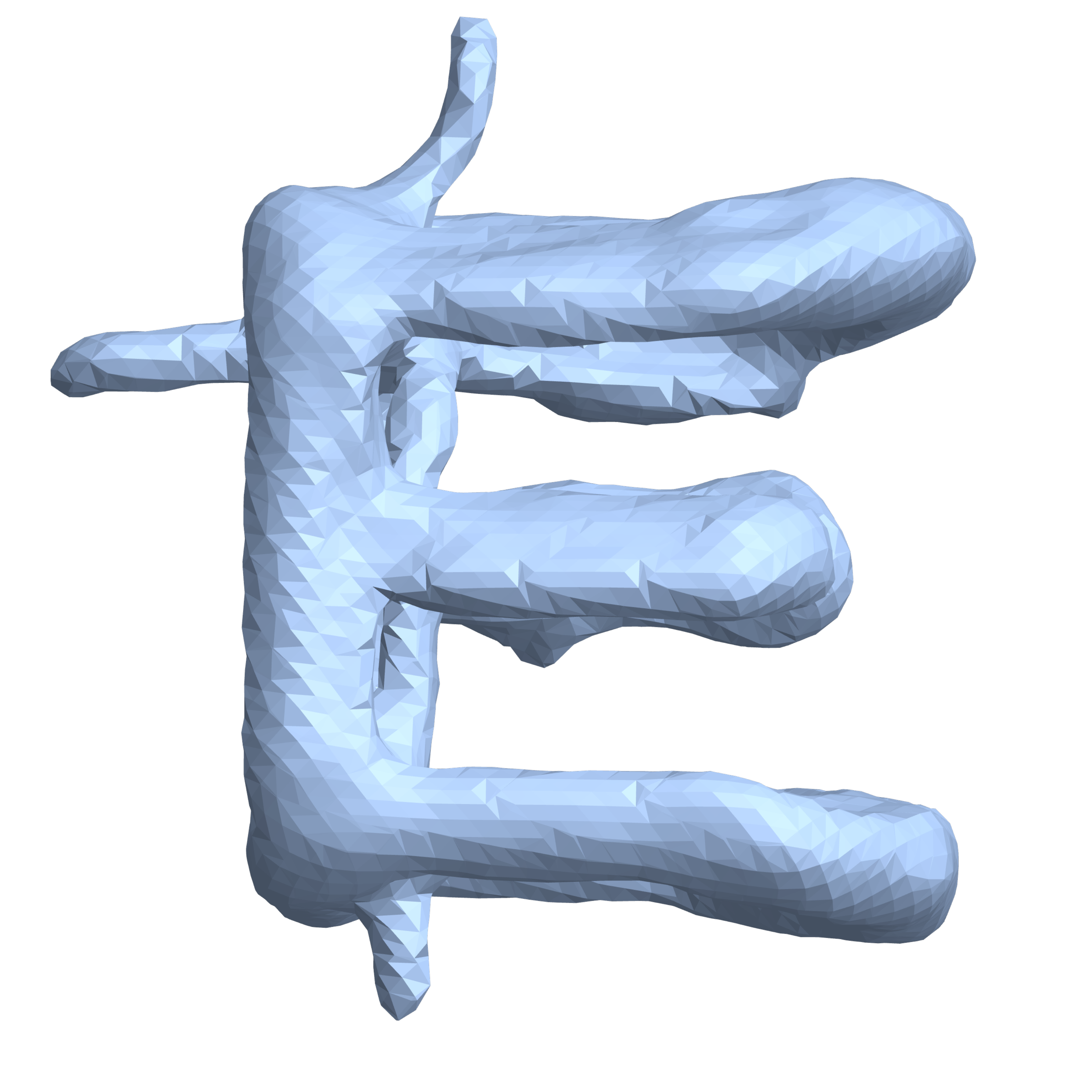}
    \end{subfigure}
    \\
    \begin{subfigure}[t]{0.32\textwidth}
        \includegraphics[width=\textwidth]{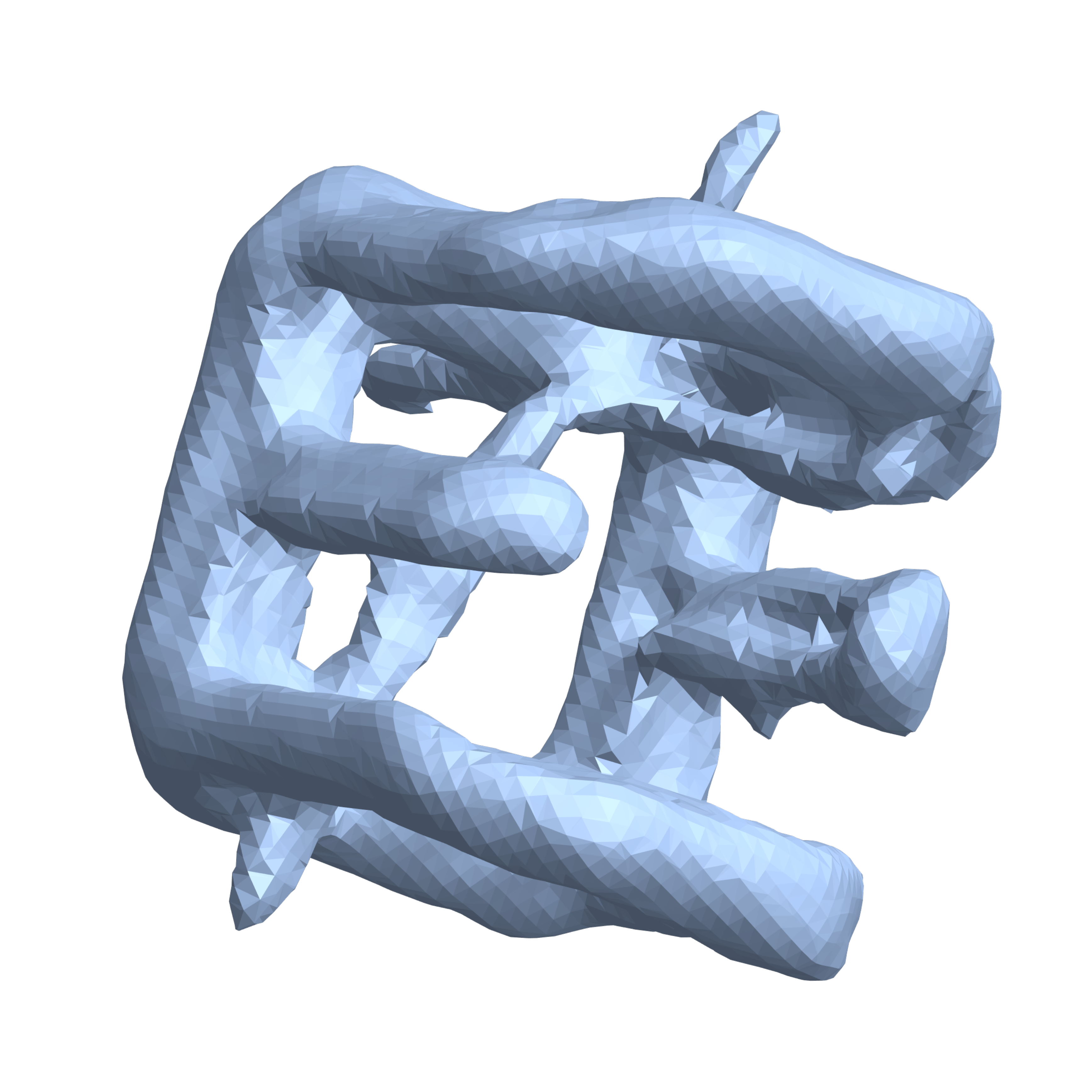}
    \end{subfigure}
    \begin{subfigure}[t]{0.32\textwidth}
        \includegraphics[width=\textwidth]{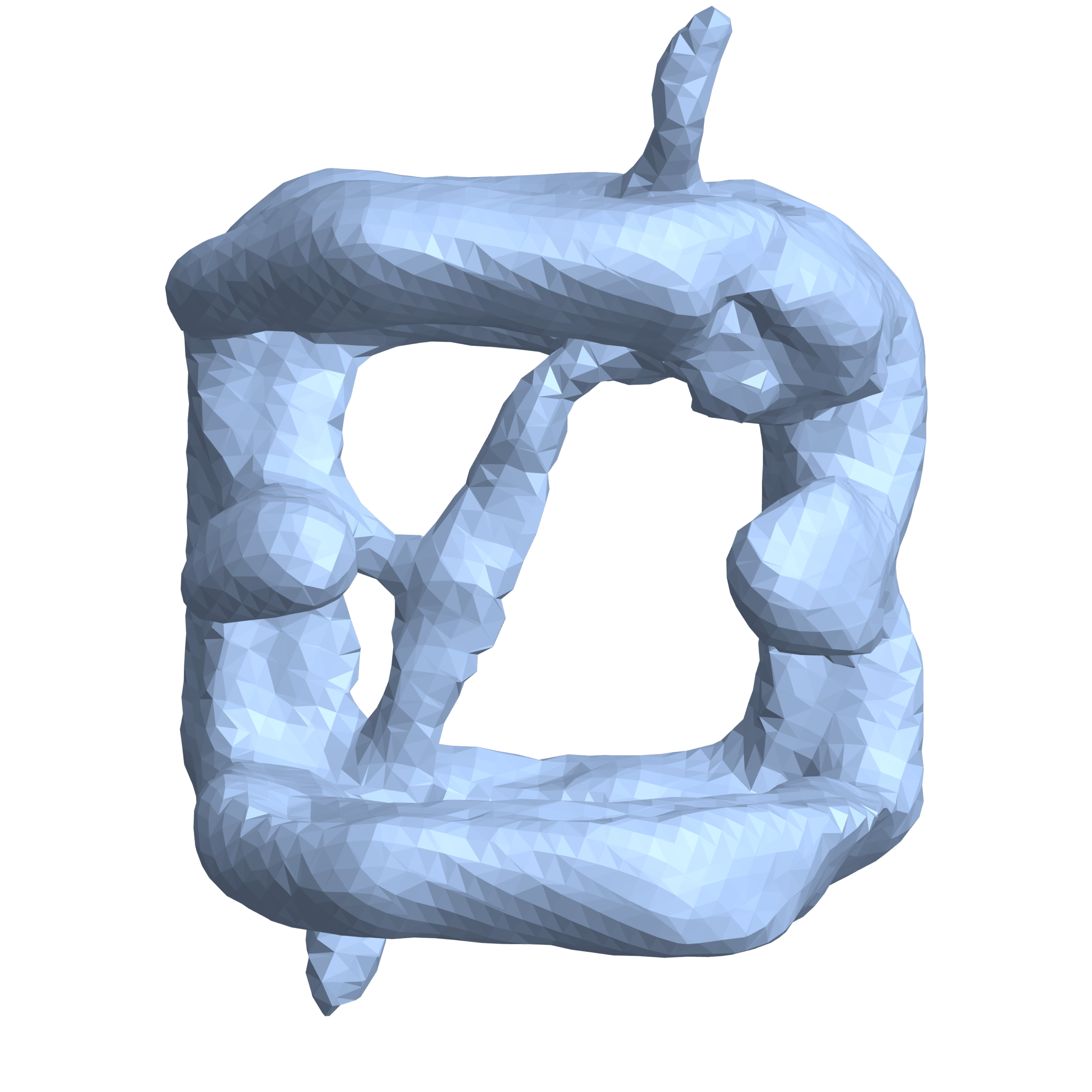}
    \end{subfigure}
    \begin{subfigure}[t]{0.32\textwidth}
        \includegraphics[width=\textwidth]{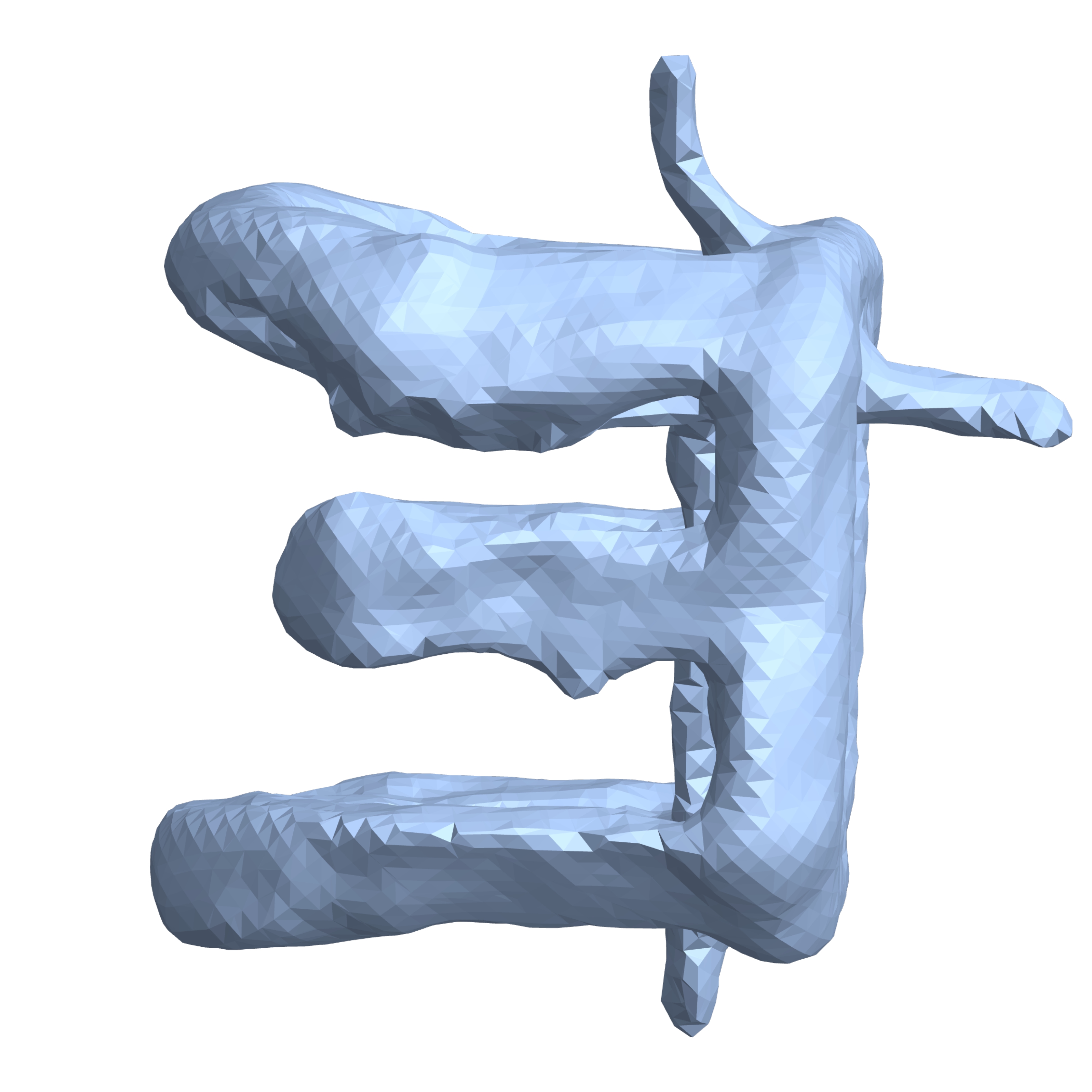}
    \end{subfigure}
    \caption{Images of reconstructed surface mesh. The top row shows similar D, O, and E views as those presented in figure~\ref{fig:DOE_cube_results}. The bottom row shows alternative views, where the first two columns show angles of the object that are not directly a D, O, or E lettering, emphasizing that we truly learn a 3D model for the imaged object, while the third column shows the reconstruction of an object face that is never directly imaged by our system (the ``back E''). While quality is worse on this face of the object, the available information from the side we do image is sufficient to infer the general structure.}
    \label{fig:DOE_mesh_views}
\end{figure}

While the regenerated training view results provide a good closure test, the NeuS model could in principle learn to reproduce the colors of these training views without learning anything about the three dimensional structure due to the flexibility offered by learning a view dependent color. As we are interested in the reconstruction of a three dimensional object, it is therefore useful to analyze the learned 3D surface, which here is the zero level set of the learned signed distance function (SDF). To do so, the learned SDF is sampled in a regular grid of $64^3$ points within a box of $\unit[2.02]{mm}$ per side centered at the nominal object position, with this sampling grid chosen to roughly match the anticipated resolution. A marching cubes algorithm within the trimesh software~\cite{trimesh} is then used to construct a triangular mesh of the learned surface. Orthographic depth maps of the three DOE letters from this learned surface are compared to the corresponding CAD model and microscope image in figure~\ref{fig:DOE_cube_results}. Figure~\ref{fig:DOE_mesh_views} directly shows the surface mesh for the three DOE letters, as well as at a few different angles, demonstrating the true 3D structure of the learned surface. The object is reconstructed with high fidelity, with very clear DOE lettering as well as more minute structure, such as the bowing out of the legs of the E. Impressively, the reconstruction also learns the $\unit[120]{\mu m}$ thread used to suspend the DOE object in front of the device, a notable artifact present relative to the CAD and microscope images. The reconstruction does suffer in regions where there is minimal information in the training views, such as the ``back E'' or the right side of the O, as shown in figure~\ref{fig:DOE_mesh_views}. However, this is expected due to the angular span of the dome, and much of the structure is still able to be inferred.

\begin{figure}[h]
    \centering
    \includegraphics[height=5in]{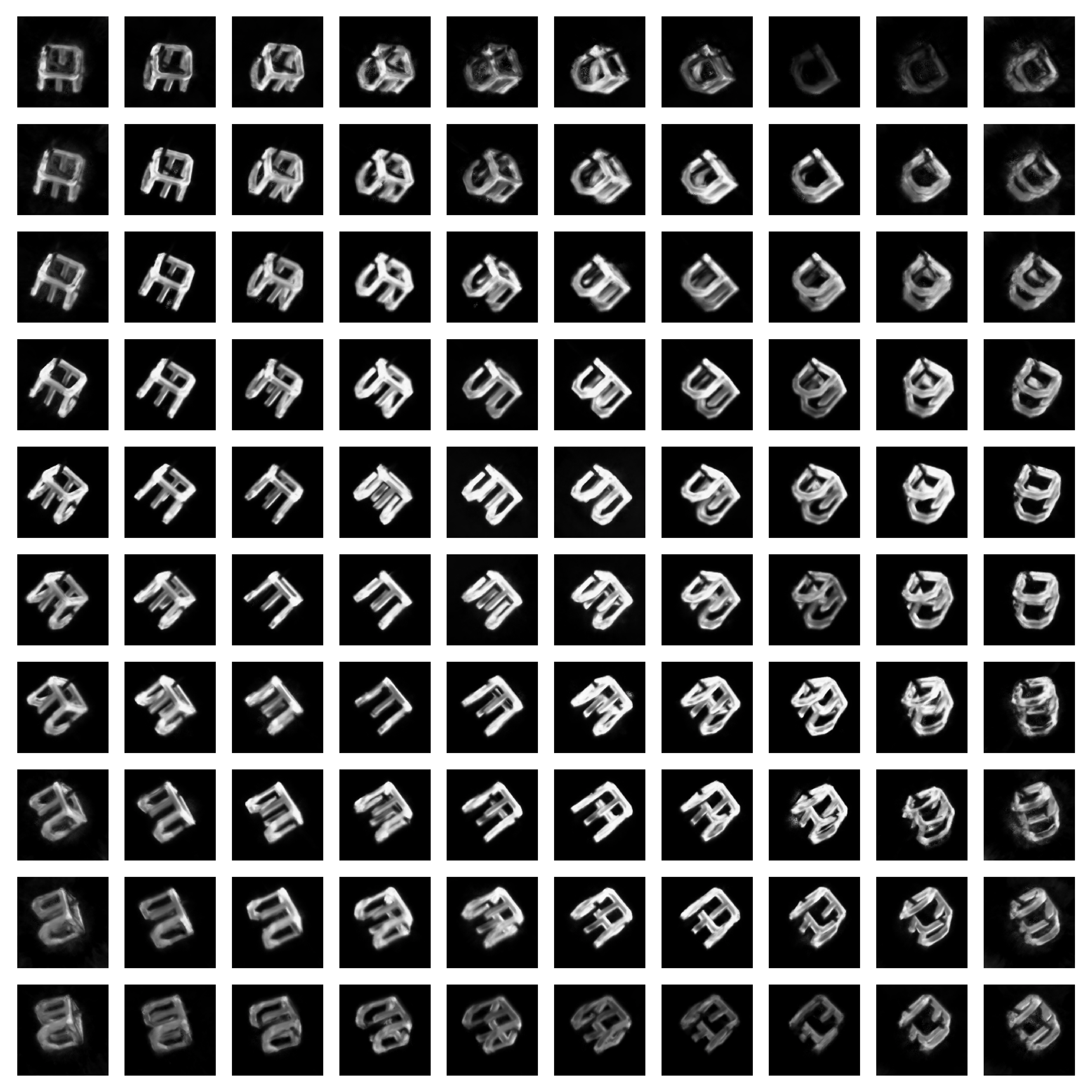}
    \caption{Interpolated views generated from the learned NeuS reconstruction. The inputs to the reconstruction are a set of rays generated at the center of the device. This set of rays is then rotated around the vertical (x-axis of the grid) and horizontal (y-axis of the grid) axes, centered on the nominal object position, and the resulting reconstruction from each set of rotated rays is shown. The range of rotation angles is between $-55$ and $55$ degrees, and resulting views contain no explicit correspondence to those used for training. Image quality is generally high, demonstrating a learned 3D structure and light field.}
    \label{fig:interp_img_grid}
\end{figure}

Information about the learned three dimensional surface is fully contained in the signed distance function, and can be visualized using colorless meshes, as done above. While such information is our primary target, it is also interesting to analyze the learned color as an insight into the learned light field, especially in our case of imperfect illumination. To do so, we sample a set of views from various positions on a sphere centered at the nominal object position. Views are sampled in an even grid of angles between $-55$ and $55$ degrees measured both vertically and horizontally from the optical axis. This set of views is shown in figure~\ref{fig:interp_img_grid}. The quality of each of these views is generally good and realistic, demonstrating a physically meaningful learned 3D structure and light field. Note that the center of the grid corresponds to the center of the device and is correspondingly the brightest, while the edges of the grid correspond to regions on the outer edge/outside of the device, and the learned intensity is correspondingly dimmer. This drop in brightness from central views to the edge views is expected since the mirrors are all of equal radii and form light limiting apertures, hence are subject to the cosine drop off discussed in section~\ref{sec:lightfield_single_view}. The effect is also exacerbated by the limited illumination used in this demonstrator.

\begin{figure}
    \centering
    \includegraphics[width=5in]{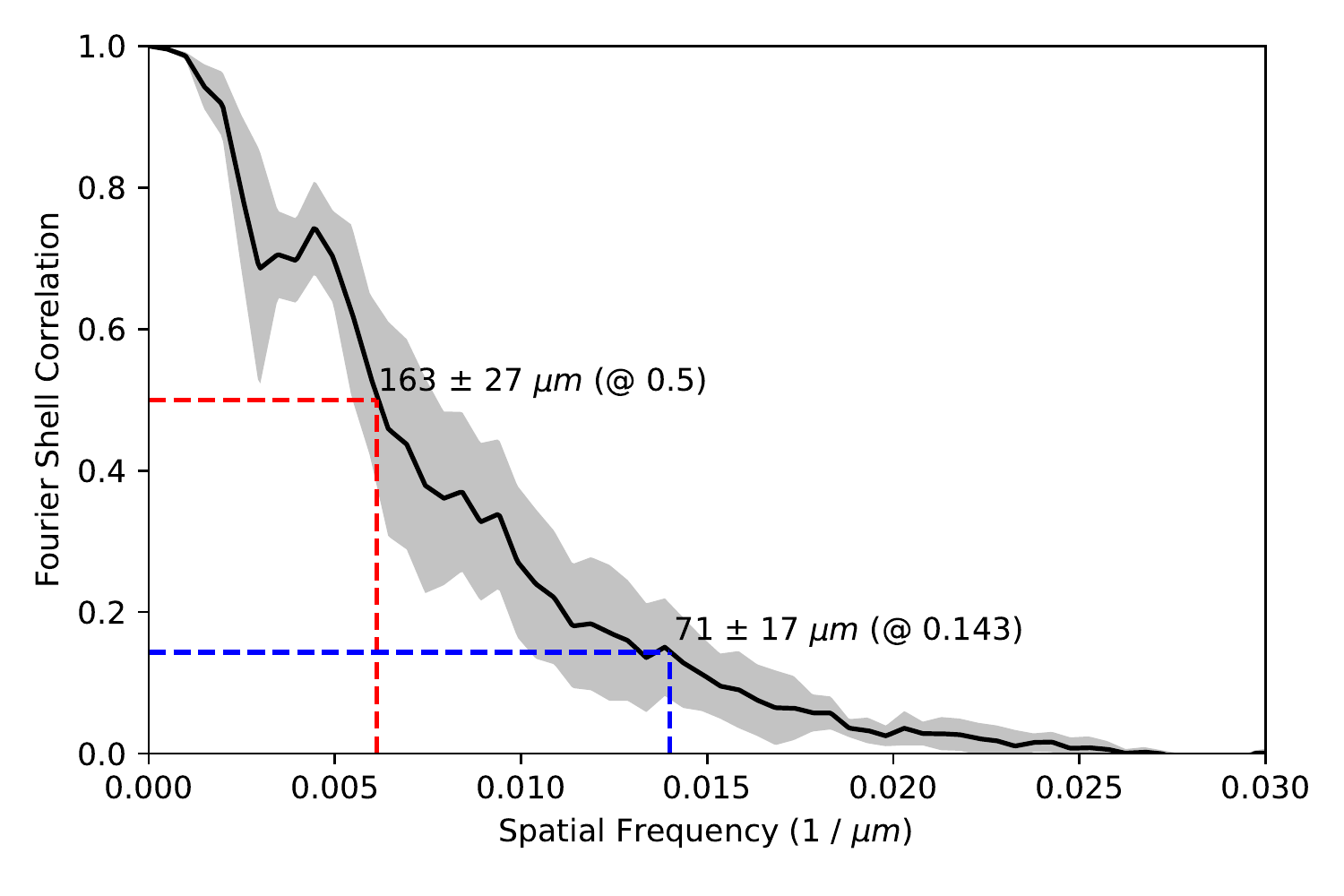}
    \caption{Fourier shell correlation curve for the reconstructed DOE object. 77 extracted views are randomly split into two distinct halves, and each half is separately reconstructed. All reconstructions are trained for 140000 iterations, and the Fourier shell correlation is calculated by comparing logistic densities of learned width centered at the learned surfaces. Results shown are the mean and standard deviation calculated from running this procedure across 5 different random splits of the 77 views. Resolutions are quoted for two commonly used thresholds, with $\unit[71\pm 17]{\mu m}$ resolution corresponding to the standard threshold used for split dataset comparisons. Results are qualitatively consistent with the size of reconstructed features.}
    \label{fig:doe-fsc}
\end{figure}

The reconstructed features shown in the above figures demonstrate the ability of our setup to reconstruct both the $\unit[142]{\mu m}$ DOE letters and the $\unit[120]{\mu m}$ thread, providing a qualitative measure of resolution for the system with the chosen reconstruction algorithm. To further quantify the resolution of the system, we use the Fourier shell correlation (FSC)~\cite{FSC}, a metric used extensively in several other contexts, including cryogenic electron microscopy, which compares the correlation between two volumes as a function of spatial frequencies. Roughly, pure signal corresponds to perfect correlation, or a FSC at 1 for all frequencies, while pure noise corresponds to no correlation, or a FSC at 0 for all frequencies. A variety of thresholds are set to determine the spatial frequency at which the transition from signal to noise occurs. As the exact orientation of the imaged object is not known, making a comparison to ground truth difficult, we follow a standard procedure of a ``split-halves'' comparison, namely, we randomly split the set of training views into two halves, run a reconstruction on each half, and calculate the correlation between these two reconstructions. Figure~\ref{fig:doe-fsc} shows the resulting Fourier shell correlation curve, with error bars derived from running this procedure with various random splits. The standard threshold for such split halves comparisons is 0.143~\cite{FSC_0143} 
, corresponding to a resolution of $\unit[71 \pm 17]{\mu m}$ for our system and the chosen dataset and reconstruction method. This resolution, as well as the regime of significant correlation shown by the full curve, is compatible with the qualitative results shown above, as well as the design goals of the demonstrator. 

\subsection{Impact of misalignment and 3D printing resolution}

\begin{figure}[h]
    \centering
    \includegraphics[height=2.5in]{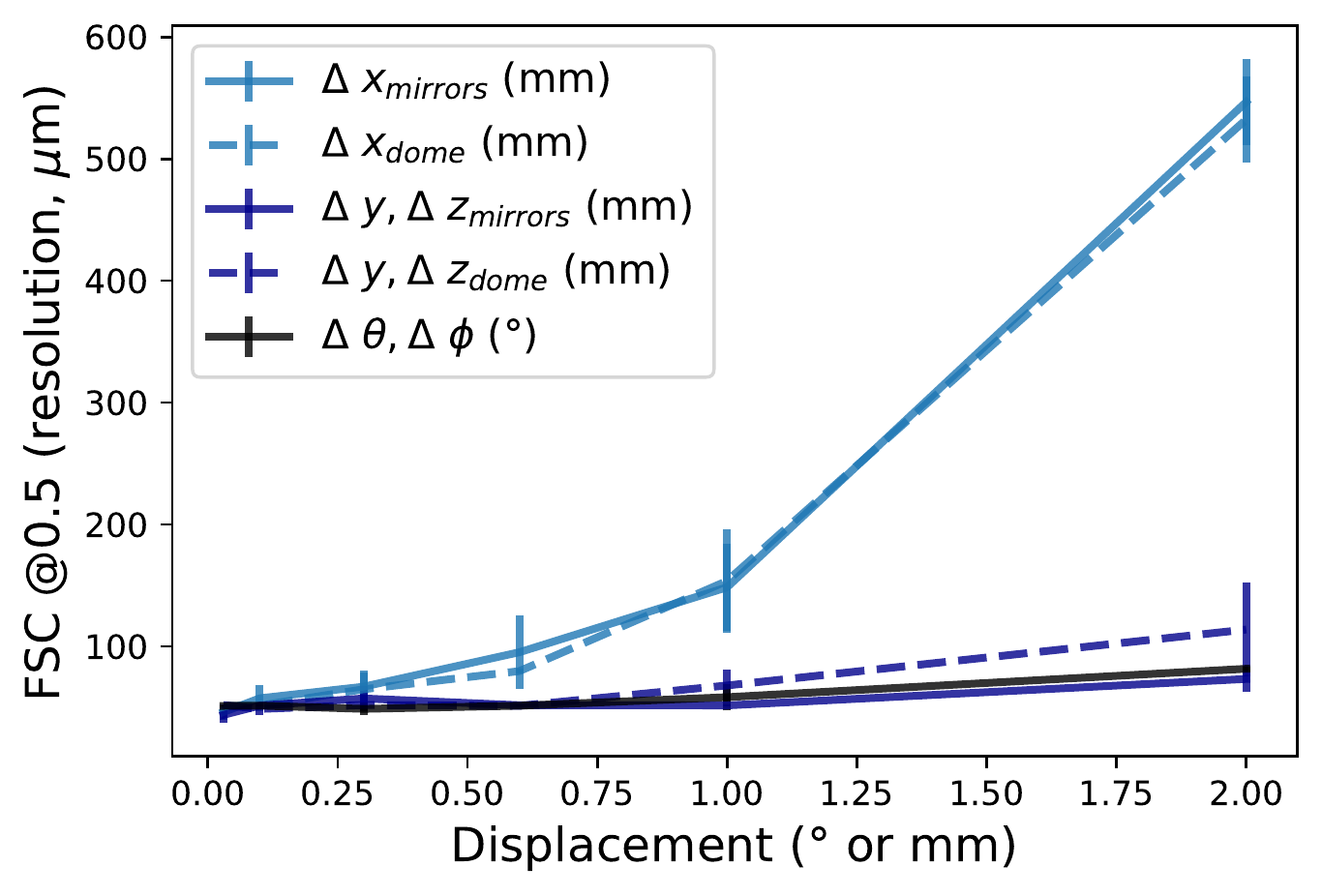}
    \caption{Resolution of the system as a function of misalignment and 3D printing error. The resolutions were obtained by computing the Fourier shell correlation between the ground truth and the 3D reconstructed models of the test object. Without calibration of the simulator, displacements below $\unit[0.6]{mm}$ or $2^{\circ}$ still allow to reconstruct $\unit[100]{\mu m}$ features. Results shown are the mean and standard deviation calculated from running the reconstruction algorithm with 5 different seeds.}
    \label{fig:fsc_misplacements}
\end{figure}

While our simulator assumes a perfectly calibrated device, the instrument is imperfect due to misalignment and 3D printing errors. Results from section~\ref{results_exp_results} suggest that in our experimental setup, those effects are small enough such that no calibration of the simulator is necessary. A detailed calibration will be explored in future work.
For better understanding the tolerances of our system with an uncalibrated simulator, we generate synthetic images of the test object with introduced displacements, and analyze how they impact the quality of the reconstruction. Five effects are considered:
\begin{itemize}
    \item Displacement of the dome along the optical axis $\Delta x_{dome}$. We simulate a displacement of the dome along the optical axis by shifting all the mirrors in the same direction.
    \item Stochastic displacement of the mirrors along the optical axis $\Delta x_{mirrors}$. We simulate 3D printing errors by shifting each mirror stochastically either in the direction along the optical axis ($x$) or in the opposite direction ($-x$).
    \item Displacement of the dome in the plane perpendicular the optical axis $\Delta y_{dome}, \Delta z_{dome}$. We simulate a displacement of the dome in the plane perpendicular the optical axis by shifting all the mirrors in the same direction.
    \item Stochastic displacement of the mirrors in the plane perpendicular the optical axis  $\Delta y_{mirrors}$, $\Delta z_{mirrors}$. We simulate 3D printing errors by shifting each mirror stochastically in different directions ($y$, $-y$, $z$ or $-z$).
    \item Stochastic displacement of the orientation of the mirrors $\Delta \theta, \Delta \phi$. We simulate 3D printing errors by changing the orientation of each mirror stochastically ($\theta$, $-\theta$, $\phi$ or $-\phi$).
\end{itemize}

For each of the described effects, we simulate an imperfect device which we use to generate a light field image of the test object. Then, we use each image to do a tomographic reconstruction with a simulator that does not account for those effects (uncalibrated), and we report the quality of the reconstruction as a function of the magnitude of the introduced displacements (see figure~\ref{fig:fsc_misplacements}). Note that we study tomographic simulation/reconstruction of the test object, differing from the assumption of an opaque test object in the previous section. This approach allows us to assess the performance of a reconstruction method closer to our target atom cloud application, but keeps the rich spectrum of features of the test object, providing a better handle for assessment of reconstruction quality. Here we report the quality in terms of the spatial resolution of the reconstructed object using the Fourier shell correlation. We report the resolution at which the Fourier shell correlation between the ground truth model and the learned one attains the commonly used threshold value of $0.5$ \cite{FSC_0143} (differing from the 0.143 threshold used for the ``split-halves'' comparison above). We observe that the displacements of the mirror orientations as well as their displacements in the plane perpendicular to the optical axis does not considerably impair the resolution of the system; $\unit[100]{\mu m}$ features can still be reconstructed with a $2^\circ$ or $\unit[2]{mm}$ deviation. On the other hand, displacements along the optical axis are much more critical and should be cautiously minimized when designing the system. 
\section{Conclusions}
\label{sec:concl}

The full exploitation of the scientific opportunities enabled by atomic sensors will require new types of imaging devices that are capable of maximizing light collection while keeping a high depth of field, and that can provide 3D imaging reconstruction capabilities. We have designed and tested a prototype of a novel light field, single-shot, imaging device suitable for the MAGIS-100 experiment. Results using a 3D-printed target demonstrate that the system achieves (and exceeds) its design requirements. 

This new type of imaging device can have broad applicability in scientific and industrial applications that can benefit from single-shot 3D imaging under low light conditions. Examples include 3D inspection of industrial parts, 3D-printing inspection lines, etc.  

Our device has several key advantages compared to other light field systems. It provides high light collection, solid angle, resolution, field of view, and depth of field, in a single-shot, and using a commercial camera and lens. It could be considered as type of 3D microscope that can fully characterize an object that may be very dim in a single shot. 
This is a fundamentally different way to do 3D microscopy than structured light microscopy that necessarily uses multiple shots or varying lighting.

\section*{Acknowledgements}

We would like to thank Andrew Rasmussen and Michael Greenberg for their invaluable contributions simulating and studying the optical system with Zemax, as well as Will Michaels for his optical measurements at the lab. We are also grateful to Gordon Wetzstein for his careful read of this manuscript and feedback provided.  
This work was supported by the Department of Energy, Laboratory Directed Research and Development program at SLAC National Accelerator Laboratory, under contract DE-AC02-76SF00515. 

\bibliographystyle{JHEP}
\bibliography{bibliography}

\newpage
\appendix
\section{Entocentric imaging - A geometric perspective}
\label{appendix:entocentric}
\begin{figure}[h]
    \centering
    \includegraphics[width=0.9\textwidth]{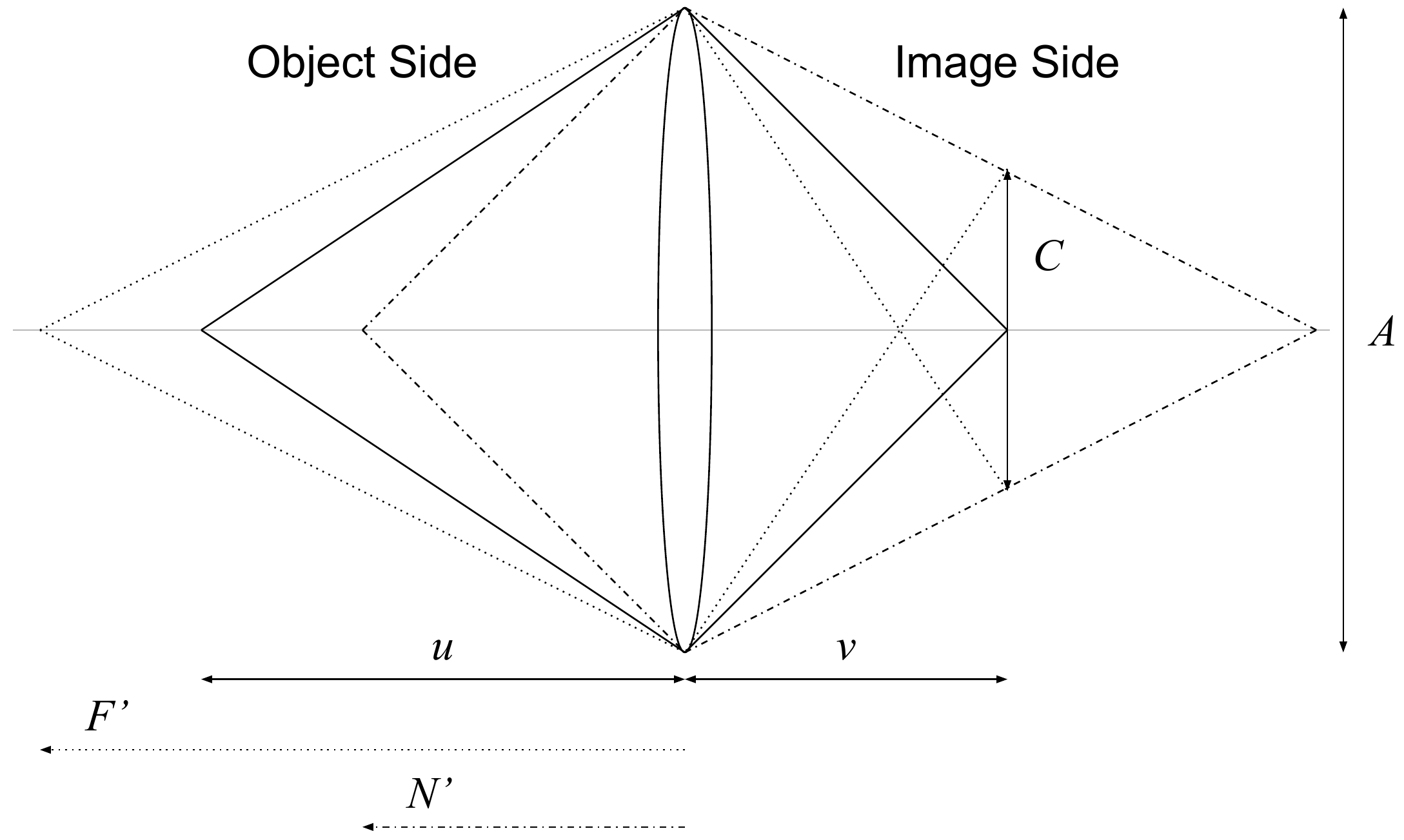}
    \caption{Ray diagram for an ideal thin converging lens of aperture diameter $A$. Only the marginal rays are shown, where each line-style corresponds to rays starting from a unique object distance along the optical axis. The rays from object distance $u$ are focused down to a point at image distance $v$. Rays starting at object distances $F'$ ($N'$) converge at a point closer (further) to the lens than $v$. In these cases, light spreads across the spot of diameter $C$ when looking at the plane at distance $v$ from the lens. This spot size corresponds to the geometric circle of confusion for the lens, and $F'-N'$ is the geometric depth of field of the lens about the nominal object distance $u$.}
    \label{fig:entocentric}
\end{figure}

Here a brief review of entocentric image formation is provided from a purely geometric point of view. Using an ideal thin converging lens as our focusing optic one can visualize the key image formation properties that are relevant to the scope of this work. Figure~\ref{fig:entocentric} shows ideal image formation for such a simple optic of focal length $f$, where the marginal rays are traced through the system for point sources located at distances $u$, $F'$, and $N'$ from the lens. A point source at the nominal object distance of $u$ from the lens is imaged by the lens to a point at a distance $v$ from the lens on the image plane side of the optic. The point at object distance $F'$ ($N'$) is focused closer (further) to the lens, hence results in a defocused image when imaged at the image distance of $v$ from the lens. 

Using the lensmaker’s equation \cite{hecht1987optics} in the thin lens paraxial approximation, the object distance $F'$ ($N'$) can be determined such that the spot size (circle of confusion) diameter at the plane of focus $v$ is equal to $C$. Note the convention here that all distances measured from the lens are positive. This depends on the lens aperture (diameter = $A$ as shown) and can be solved for using the similarity of triangles, resulting in:
\begin{equation*}
    F' = \frac{fuA}{fA - C(u-f)}, \quad N' = \frac{fuA}{fA + C(u-f)}
\end{equation*}
and
\begin{align}
    \label{eqn:ento_dof}
    DoF = F' - N'
    &= \frac{2fuAC (u - f) }{(fA)^2 - C^2 (u - f)^2}  \\
    \label{eqn:ento_dof_simplified}
    &\approx 2CN \left( 1 + \frac{1}{m} \right)^2
\end{align}
where $DoF$ is the geometric depth of field is the range of object distances over which the circle of confusion resulting from the geometric defocus of the lens is less than some value $C$. The final approximation is made in the limit that the circle of confusion $C$ is significantly less than $u$, $f$, and $A$, and using the lens $f$-number $N = f/A$ and the lens magnification $m = f/(u-f)$ where positive $m$ corresponds to an inverted image. In this limit, $DoF$ is independent of the lens focal length and can be interpreted as function of $N$ and $m$ for a given $C$. 

This simple model can also provide a geometric estimate of the total light collected from an isotropically fluorescing point source that is being focused by the lens. Assuming that the lens shown in figure~\ref{fig:entocentric} has uniform light collection over its aperture of diameter $A$, the fraction of light collected by the lens is (also a function of $N$ and $m$),
\begin{align}
\label{eqn:ento_light}
\frac{1}{4\pi} \times 2\pi \left(1 - \cos{\left(2\arctan{\frac{A}{2u}}\right)}\right) &= \frac{1}{2} \left(1 - \cos{\left(2\arctan{\frac{m}{2N(1+m)}}\right)}\right)
\end{align}

\clearpage
\section{Normal angles for a folded view}
\label{appendix:normal_equations}

\begin{figure}[h]
    \centering
    \includegraphics[width=0.9\textwidth]{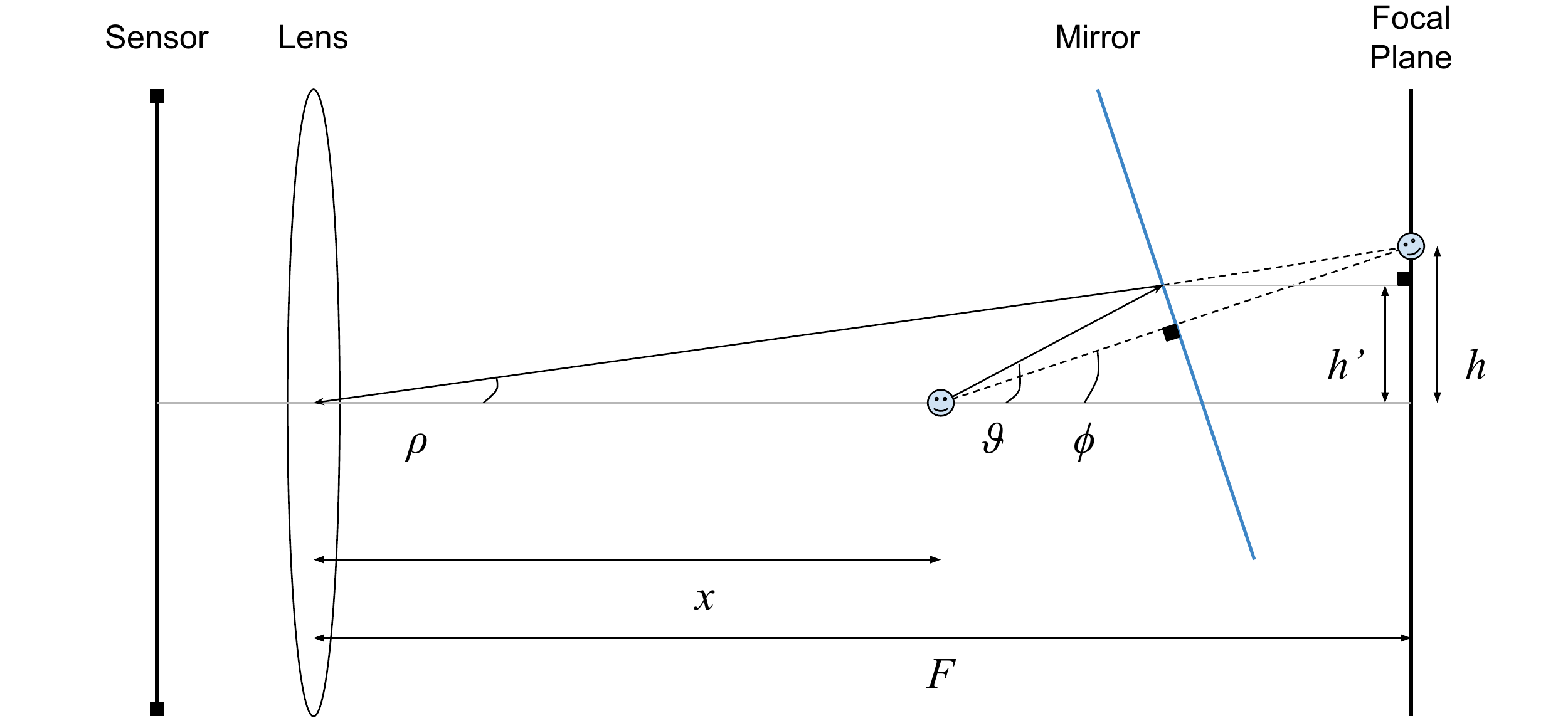}
    \caption{Principal ray diagram of a single folded view. Here the object (smiley face) is placed at a distance $x$ from the lens center, offset from the focal plane of the lens which is at a distance $F$ from the lens. A flat mirror is introduced between the object and the focal plane, which produces a virtual object that lies on the focal plane; this virtual object is focused onto the sensor by the lens. The mirror normal angle $\phi$ is chosen such that the principal (central) ray emanating from the object (at angle $\theta$) is reflected towards the center of the lens. This constrains the geometry of the mirror and determines the mirror normal angle as described in appendix~\ref{appendix:normal_equations}. Note that the $\theta>0$ view angle results in an image that sees the object as if it were rotated by angle $\theta$ about the optical axis. Note also that $\rho = 2\phi - \theta$}
    \label{fig:singleview_folded}
\end{figure}

Figure~\ref{fig:singleview_folded} is used as the reference for the following equations. We start by requiring that the height of the point of reflection $h’$ is consistent when derived from the lens center perspective and from the object’s perspective. Note that $r = F/x$ is used for simplification, along with the fact that $\rho = 2\phi-\theta$ which is shown geometrically in figure~\ref{fig:singleview_folded}.
\begin{align}
\frac{\frac{F-x}{2\cos{\phi}}\sin{\theta}}{\cos{(\theta-\phi)}} &= h’ \nonumber\\
h’ &= x\left[ \frac{1}{\tan{\rho}} - \frac{1}{\tan{\theta}} \right]^{-1} \nonumber\\
(r-1)\tan{\theta} &= \sin{\rho}\left[ \frac{r}{\cos{\rho}} + \frac{1}{\cos{\theta}} \right] \label{eqn:normal_derivation}
\end{align}
This equation is solved numerically for $\cos{\rho}$ for specific values of $r$ and $\theta$. Note that solving for $\rho$ fully constrains the mirror parameters $\phi$ and $h'$.

\begin{figure}
    \centering
    \includegraphics[width=0.8\textwidth]{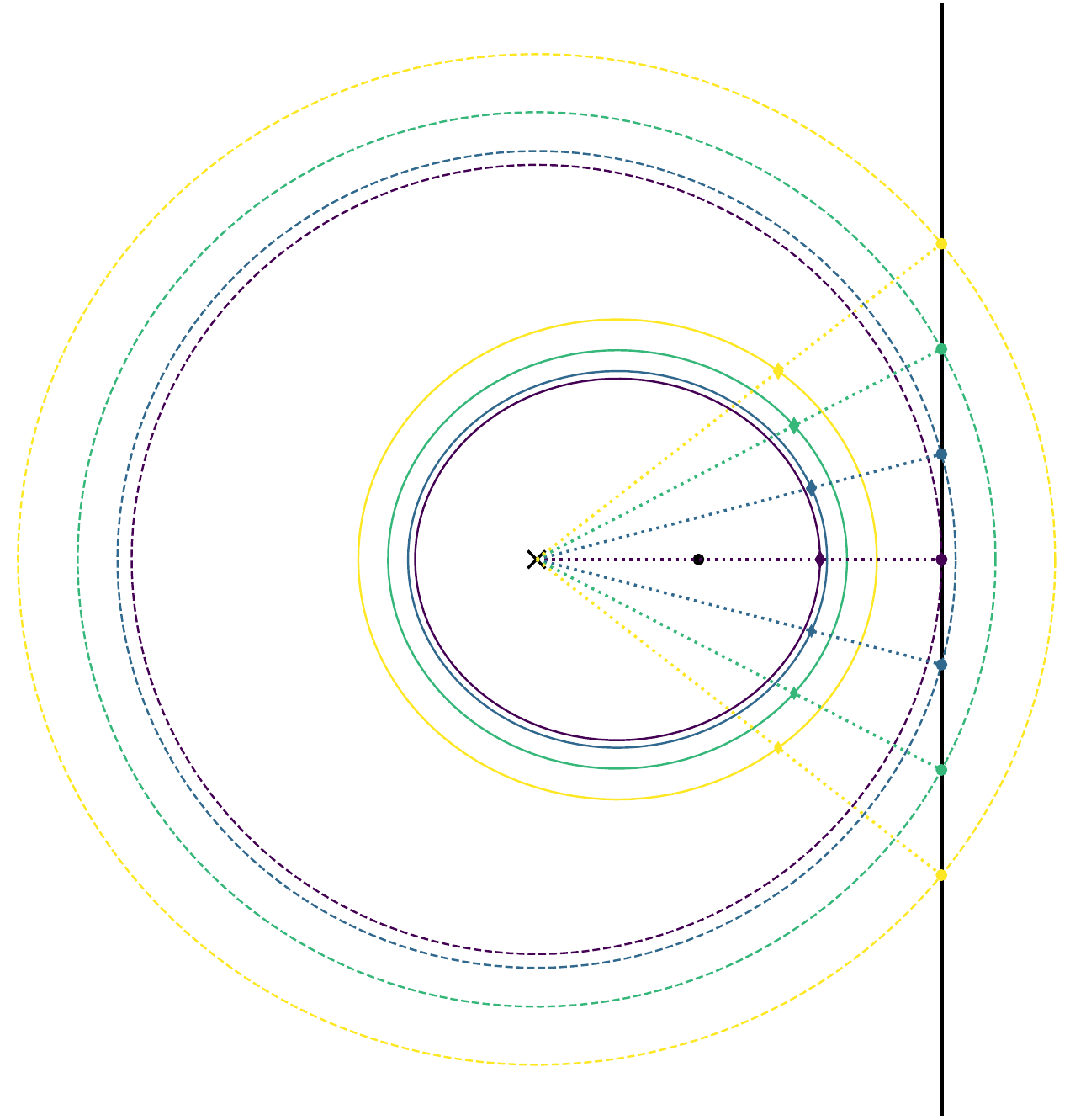}
    \caption{Image formation geometry for a single folded view like figure~\ref{fig:singleview_folded}. Here the black X marks the location of the lens center and the black dot the location of the object. The thick solid black vertical line on the right corresponds to the plane at which the lens is focused. The solid colored ellipses represent the family of ellipses that have the lens center and object as the two focii. The dashed circles are the corresponding circular directrices of the ellipses, centered at the lens center. The colored diamonds show the mirror center positions and the colored dots show the corresponding virtual object locations at the plane of focus. The dotted lines connect the lens center, mirror centers, and virtual object locations.}
    \label{fig:multiview_ellipse}
\end{figure}

Alternatively, the system can be geometrically solved using the properties of ellipses. Figure~\ref{fig:multiview_ellipse} shows the construction that be used to find both the mirror normal angles and the mirror center positions. For any point on the plane of focus where a virtual image of the object is to be formed, a unique circular directrix from the family of ellipses will pass through that point. The point of intersection of the line between the lens center and the point on the plane of focus, with the ellipse identifies the location of the corresponding mirror center. The normal to the ellipse at the intersection point corresponds to the normal angle of the mirror. This construction also allows the determination of the mirror parameters given a non-flat plane of focus, since the procedure can be identically applied in that case.

\clearpage
\section{Field of view for a folded view}
\label{appendix:fov_equations}

Figure~\ref{fig:singleview_folded_fov} is used as the reference for the following equations. We start by finding the mirror radius projected onto the perpendicular to the central ray direction, $l$. Then, we find the distance between the lens center and the center of the mirror, $D$. These two quantities are then combined to determined $\chi$, which is the used to determine the field of view, $FoV$.
\begin{align}
l &= R\cos{(\theta-\phi)}
\nonumber \\
D &= \frac{h'}{\sin{\rho}} = \frac{(F-x)\sin{\theta}}{2\cos{\phi}\sin{\rho}\cos{(\theta-\phi)}}
\nonumber \\
\tan{\chi} &= \frac{l}{D} = \frac{2R\cos^2{(\theta-\phi)}\cos{\phi}\sin{\rho}}{(F-x)\sin{\theta}}
\nonumber \\
FoV &= F\left[ \tan{(\rho + \chi)} - \tan{(\rho - \chi)} \right]\label{eqn:fov}
\end{align}

\begin{figure}[h]
    \centering
    \includegraphics[width=0.7\textwidth]{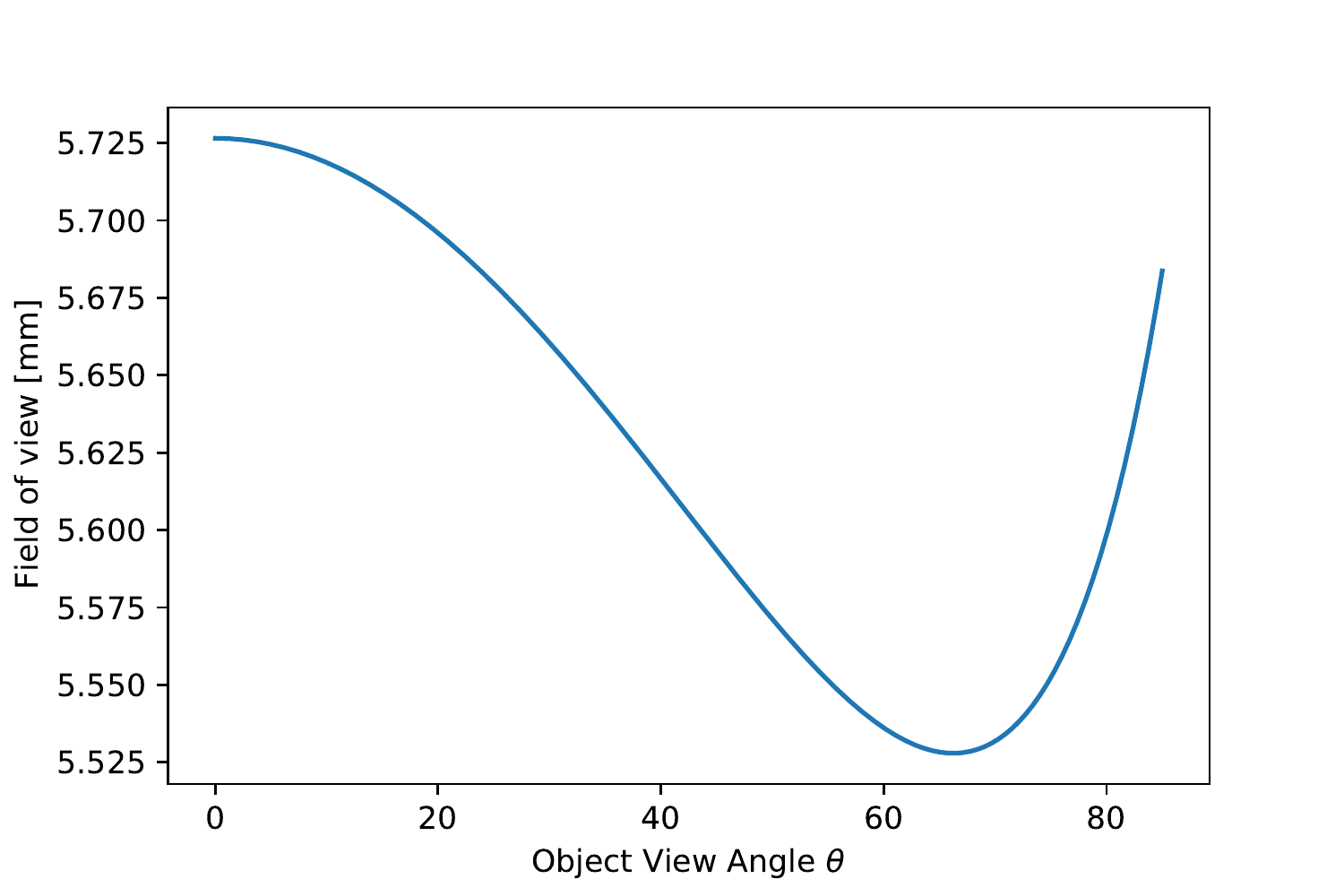}
    \caption{Field of view as a function of the view angle $\theta$ determined by solving eq.~\ref{eqn:fov}. For this plot, an ideal lens of focal length $\unit[62.275]{mm}$ operating at a magnification of $m=0.215$ is chosen, where the object distance is picked to allow the $55^\circ$ view to lie at the edge of a full frame sensor. This is the same setup as described in section~\ref{sec:lightfield_multi_view} and the object distance is picked following eq.~\ref{eqn:object_distance}. The mirror here is a flat mirror of diameter $\unit[5]{mm}$ centered appropriately to satisfy the constraints on the principal rays.}
    \label{fig:dof_example}
\end{figure}

The variation of the field of view with the view angle $\theta$ for a particular set of parameters is shown in figure~\ref{fig:dof_example}. There are two competing effects that define the field of view of the system. Firstly, the mirror gets increasingly inclined towards the optical axis as view angle gets larger, and this results in a reduced field of view footprint due to a $\cos{(\theta-\phi)}$ factor. Secondly, the mirrors move further away from the axis as the view angler gets larger to satisfy the constraint for the virtual objects to land on the focal plane - this results in a larger field of view footprint. Therefore, the field of view variation as shown in figure~\ref{fig:dof_example} depicts a minima and a diverging field of view as $\theta$ approaches $90^\circ$. The mirror sizes can be chosen based on the field of view to account for the object of interest. Hence, the system is flexible in its ability to accommodate objects of various scales.

\clearpage
\section{Generated Image Grids}
\label{appendix:image_grids}

\begin{figure}[h]
    \centering
    \includegraphics[height=5in]{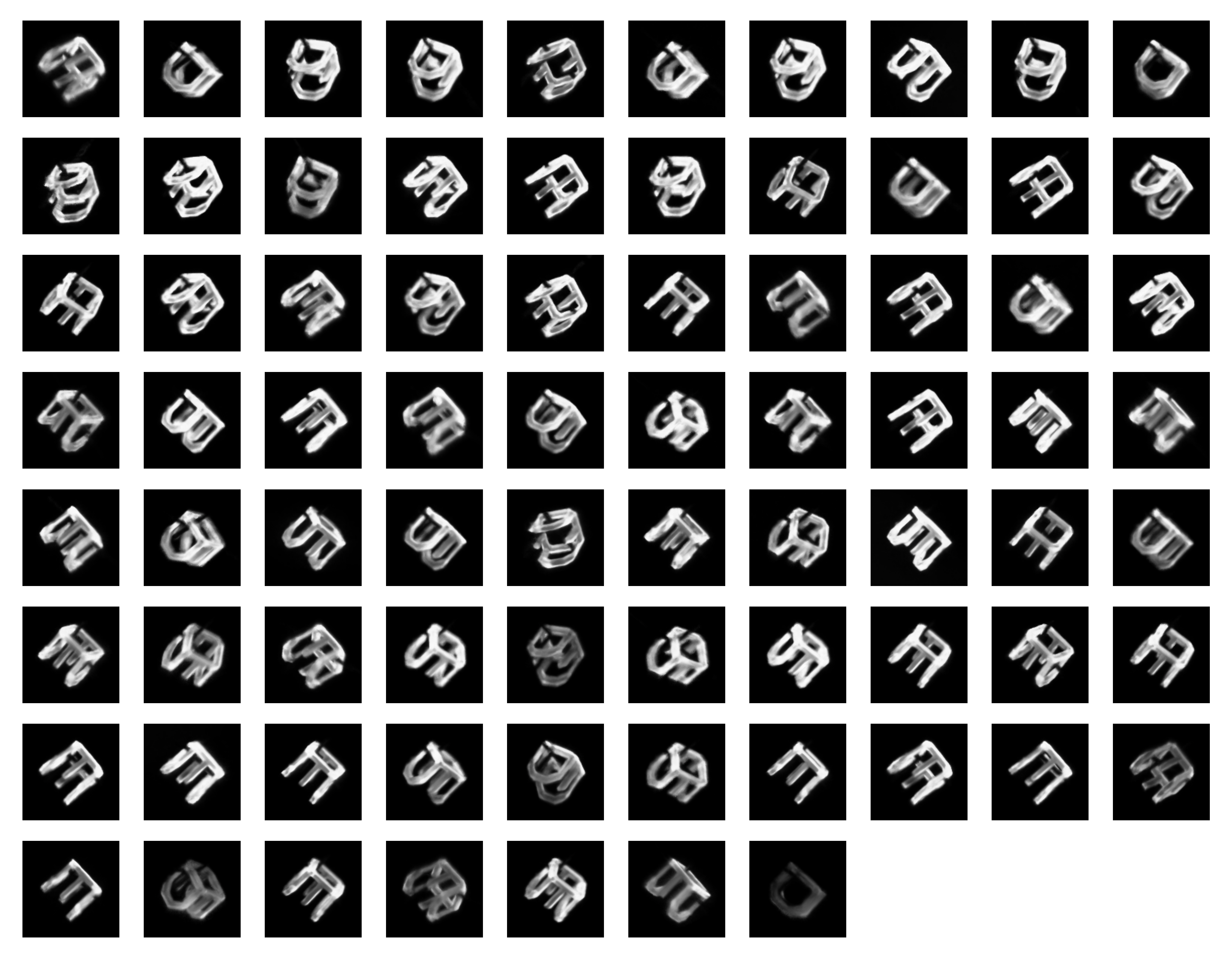}
    \caption{Views from the learned NeuS reconstruction corresponding to those used in training. Relative to the captured images shown in figure ~\ref{fig:real_img_grid}, results are visually nearly identical, demonstrating reasonable closure of the model.}
    \label{fig:gen_img_grid}
\end{figure}

Figure~\ref{fig:gen_img_grid} shows the full grid of 77 views generated from the NeuS model, corresponding to the 77 views used for training shown in figure~\ref{fig:real_img_grid}. As discussed in figure~\ref{fig:real_gen_compare}, the real and generated views are visually nearly identical, demonstrating reasonable closure of the model.

\begin{figure}[h]
    \centering
    \includegraphics[height=5in]{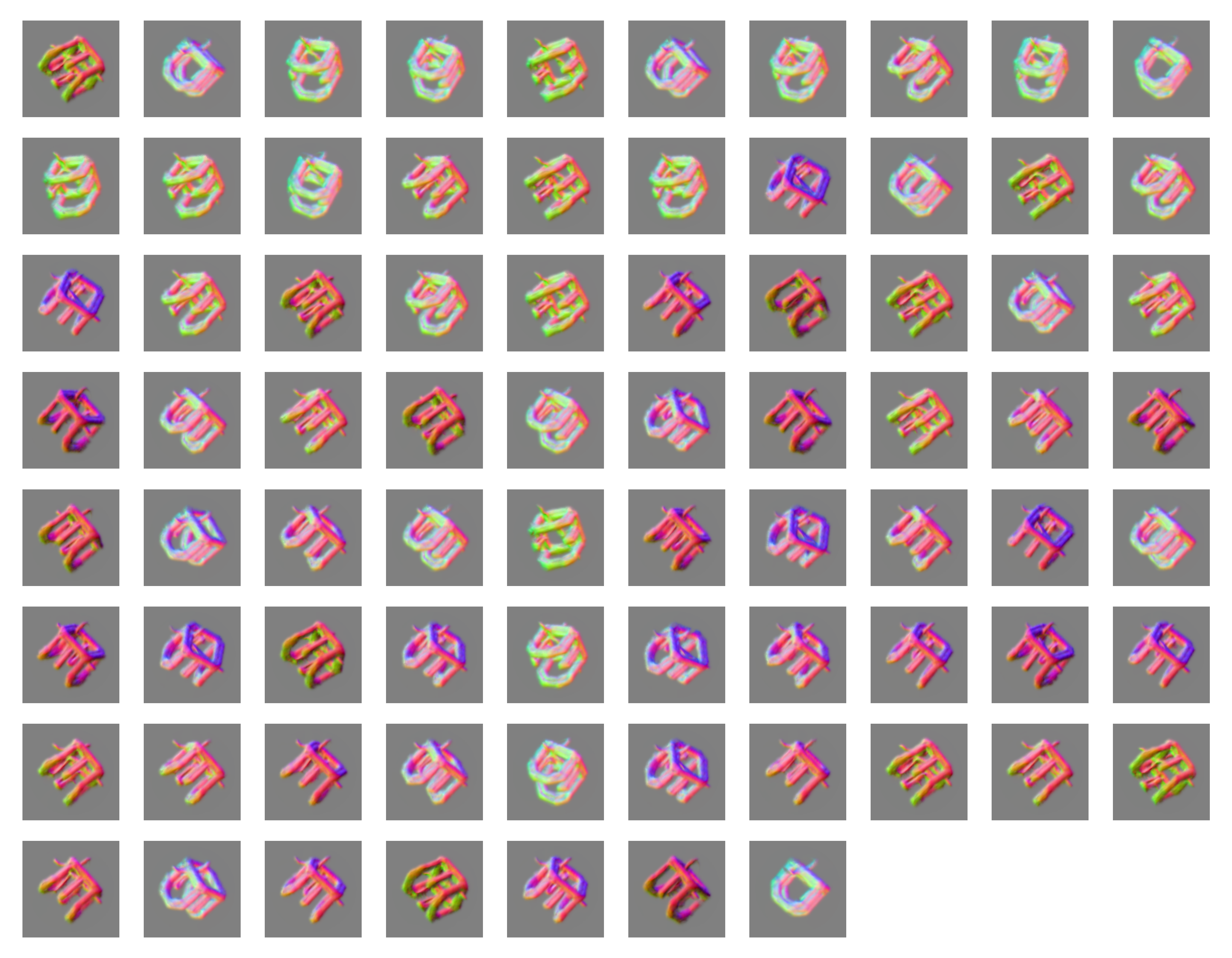}
    \caption{Generated grid of surface normals from the learned NeuS model, corresponding to the views used in training (and the generated images of figure~\ref{fig:gen_img_grid}). The colors correspond to the magnitude of the surface normal vectors when projected onto the corresponding 2D image plane. The learned 3D structure of the DOE object is apparent, as expressed via the meshes and depth maps of figures~\ref{fig:DOE_cube_results} and ~\ref{fig:DOE_mesh_views}.}
    \label{fig:gen_img_grid_normals}
\end{figure}

Figure~\ref{fig:gen_img_grid_normals} shows a grid of generated images corresponding to the same 77 training views, with information about the normal vectors of the learned surface. Colors correspond to the magnitude of these normal vectors when projected into the 2D image plane for each view. The learned 3D structure is evident, with clearly learned letter and thread structure, as already shown by figures~\ref{fig:DOE_cube_results} and ~\ref{fig:DOE_mesh_views}. Note that there are sometimes small shifts in apparent location of the colored objects of figure~\ref{fig:gen_img_grid} relative to the projections of the surface in figure~\ref{fig:gen_img_grid_normals}, demonstrating the way that NeuS deals with misalignments -- the learned surface aggregates the consistent 3D structure across all of the views, and remaining misalignments for individual views are handled via coloration of that learned surface.

\clearpage
\section{Implementation Details}
\label{appendix:neus_params}
The architecture and parameters used for the NeuS reconstruction generally follow those used for the results presented in ref.~\cite{NeuS}, described there as the `w/o mask' setting (see NeuS Appendix D). For completeness, we include the corresponding settings here, noting that the description closely follows the corresponding appendix in ref.~\cite{NeuS}, though we have highlighted areas where our implementation differs. Further note that the implementation of the networks and training procedures is taken directly from the NeuS GitHub repository.

We use two multi-layer perceptrons (MLPs) to encode the signed distance function (SDF) and color (or, here, intensity). The SDF is modeled by an MLP with 8 hidden layers of 256 nodes each. Softplus activation functions with $\beta = 100$ are used as the activation functions for all hidden layers. A skip connection connects the input with the output of the fourth layer.

The function for color prediction is modeled by an MLP with 4 hidden layers of 256 nodes each. Though only a single intensity channel is relevant for the captured images here, we convert this intensity into RGB values to match the implementation from NeuS. The color function takes as input the spatial location $\mathbf{p}$, the view direction $\mathbf{v}$, the normal vector of the SDF, $\mathbf{n}$ and a 256-dimensional feature vector from the SDF MLP. Positional encoding is applied to both spatial location $\mathbf{p}$ and view direction $\mathbf{v}$, with 6 and 4 frequencies respectively. Weight normalization is used to stabilize the training process.

Rays used as inputs to the NeuS rendering are generated from custom simulation code reproducing the experimental setup. Design parameters are used for this setup: an $f/1.2$ lens of focal length $\unit[62.275]{mm}$, focused for magnification $m=0.215$, and a camera with pixel size $\unit[3.76]{\mu m}$. In addition to these optical parameters, a scaling and centering is performed so that the NeuS rendering is done in a unit sphere centered at the origin. In our case, this means shifting and scaling the rays generated from the simulated system. Here, the ray coordinates are scaled by a factor of 1000 and shifted by the nominal object position. This corresponds to a reconstruction in a sphere of radius $\unit[1]{mm}$ centered at the nominal object position.

Networks are trained using the Adam optimizer, with a learning rate that is linearly warmed up from 0 to $5 \times 10^{-4}$ in the first 5000 iterations, and then controlled by a cosine decay schedule to a minimum of $2.5 \times 10^{-5}$. We train our baseline model for 100k iterations, which we note is shorter than the training time used by NeuS. We found that this was sufficient for our dataset, with longer training times offering negligible improvement in the quality of the reconstructed object.

A hierarchical sampling is used for the integration along each ray. This sampling starts with a uniformly sampled 64 points along the ray, and importance sampling is iteratively conducted four times, adding 16 points each time, for a total of 128 sampled points. An additional 32 points are sampled outside of the bounding sphere. 

\end{document}